\documentclass[nonacm,balance=false,sigconf]{acmart}

\usepackage{listings}
\usepackage{multirow}
\usepackage[]{xcolor}
\usepackage[T1]{fontenc}
\usepackage{graphicx} %
\usepackage{xspace}
\usepackage[nobreak]{mdframed}
\usepackage{pgfmath} % Provides math functions
\usepackage[edges]{forest}
\usepackage{paralist}
\usepackage{hyperref}
\usepackage[english]{babel}
\addto\extrasenglish{
}
\usepackage{pifont}
\usepackage{float}
\usepackage{xurl}
\usepackage{cleveref}% Load AFTER hyperref!
\usepackage{tablefootnote}
\usepackage{graphicx}
\usepackage{booktabs}
\usepackage{color}
\usepackage{tabularray}
\usepackage[htt]{hyphenat}
\usepackage{amsmath}
\usepackage{natbib}
\usepackage{balance}

\newcommand{\TODO}[1]{\textcolor{red}{#1}\GenericWarning{}{LaTeX Warning: TODO: #1}}\newcommand\todo\TODO
\newcommand{\newc}[1]{\textcolor{black}{#1}}

\newcommand{\jNorm}{\textsc{jNorm}\xspace}
\newcommand{\ossrebuild}{\textsc{OSS-Rebuild}\xspace}
\newcommand{\chainsrebuild}{\textsc{Chains-Rebuild}\xspace}
\newcommand{\diffoscope}{\textsc{diffoscope}\xspace}
\newcommand{\cmark}{\ding{51}}%
\newcommand{\unreproduciblereleases}{\newc{\textsc{8,292}}\xspace}
\newcommand{\unreproducibleartifacts}{\newc{\textsc{12,803}}\xspace}
\newcommand{\unreproducibleartifactswithjvmbytecode}{\newc{\textsc{936}}\xspace}
\newcommand{\totalartifactsinallunreproduciblereleases}{\textsc{35,956}\xspace}
\newcommand{\jnormsuccess}{\textsc{275}\xspace}
\newcommand{\jnormfailure}{\textsc{508}\xspace}
\newcommand{\jnormerror}{\textsc{153}\xspace}
\newcommand{\canonicalizedreleases}{\textsc{2,493}\xspace}
\newcommand{\ossrebuildsuccess}{\textsc{1,205}\xspace}
\newcommand{\ossrebuildfailure}{\textsc{11,598}\xspace}
\newcommand{\chainsrebuildsuccess}{\textsc{3,405}\xspace}
\newcommand{\chainsrebuildfailure}{\textsc{9,398}\xspace}

\newcommand{\GHilight}{\makebox[0pt][l]{\color{green}\rule[-4pt]{\linewidth}{10pt}}}
\newcommand{\RHilight}{\makebox[0pt][l]{\color{pink}\rule[-4pt]{\linewidth}{10pt}}}

\newcommand{\rqtaxonomy}{What are the causes of unreproducible builds in Java?}
\newcommand{\rqjnorm}{To what extent does bytecode canonicalization with jNorm mitigate unreproducible builds in Java?}
\newcommand{\rqossrebuild}{To what extent does artifact canonicalization using OSS-Rebuild make a Maven release reproducible?}

\definecolor{javared}{rgb}{0.6,0,0} % for strings
\definecolor{javagreen}{rgb}{0.25,0.5,0.35} % comments
\definecolor{javapurple}{rgb}{0.5,0,0.35} % keywords
\definecolor{javadocblue}{rgb}{0.25,0.35,0.75} % javadoc

\lstdefinestyle{customstyle}{
  backgroundcolor=\color{gray!20}, % Set the background color
  basicstyle=\ttfamily\scriptsize, % Use a monospaced font
  frame=tb, % Add a frame around the listing
  captionpos=b, % Position the caption at the bottom
  breaklines=true, % Enable line breaking
  numbers=left, % Add line numbers on the left
  numberstyle=\tiny\color{gray}, % Style for line numbers
  commentstyle=\color{black!40}, % Style for comments
  morecomment=[l]{//}, % Define // as comment delimiter,
}

\lstdefinestyle{diff}{
    escapechar=\%,
    gobble=4,
}

\begin{document}

% \title{Repair of Unreproducible Builds in Java via Canonicalization}
%\title{Mitigation of Unreproducible Builds in Java via Canonicalization}
\title{\newc{Causes and Canonicalization of Unreproducible Builds in Java}}

\author{Aman Sharma}
\affiliation{
   \institution{KTH Royal Institute of Technology}
   \city{Stockholm}
   \country{Sweden}
}
\email{amansha@kth.se}

\author{Benoit Baudry}
\affiliation{
    \institution{Université de Montréal}
    \city{Montréal}
    \country{Canada}
}
\email{benoit.baudry@umontreal.ca}

\author{Martin Monperrus}
\affiliation{
    \institution{KTH Royal Institute of Technology}
    \city{Stockholm}
    \country{Sweden}
}
\email{monperrus@kth.se}

\begin{abstract}
  The increasing complexity of software supply chains and the rise of supply chain attacks have elevated concerns around software integrity.
  Users and stakeholders face significant challenges in validating that a given software artifact corresponds to its declared source.
  Reproducible Builds address this challenge by ensuring that independently performed builds from identical source code produce identical binaries.
  However, achieving reproducibility at scale remains difficult, especially in Java, due to a range of non-deterministic factors and caveats in the build process. In this work, we focus on reproducibility in Java-based software, archetypal of enterprise applications.
  We introduce a conceptual framework for reproducible builds, we analyze a large dataset from Reproducible Central, and we develop a novel taxonomy of six root causes of unreproducibility.
  We study actionable mitigations: artifact and bytecode canonicalization using \ossrebuild and \jNorm respectively.
  \newc{Finally, we present \chainsrebuild (improvements to \ossrebuild), a tool that raises reproducibility success from 9.48\% to 26.60\% on  \unreproducibleartifacts unreproducible artifacts.}
  To sum up, our contributions are the first large-scale taxonomy of build unreproducibility causes in Java, a publicly available dataset of unreproducible builds, and \chainsrebuild, a canonicalization tool for mitigating unreproducible builds in Java.
\end{abstract}


\maketitle
\pagestyle{plain}

\begin{figure*}
  \centering
  \includegraphics[width=0.8\textwidth]{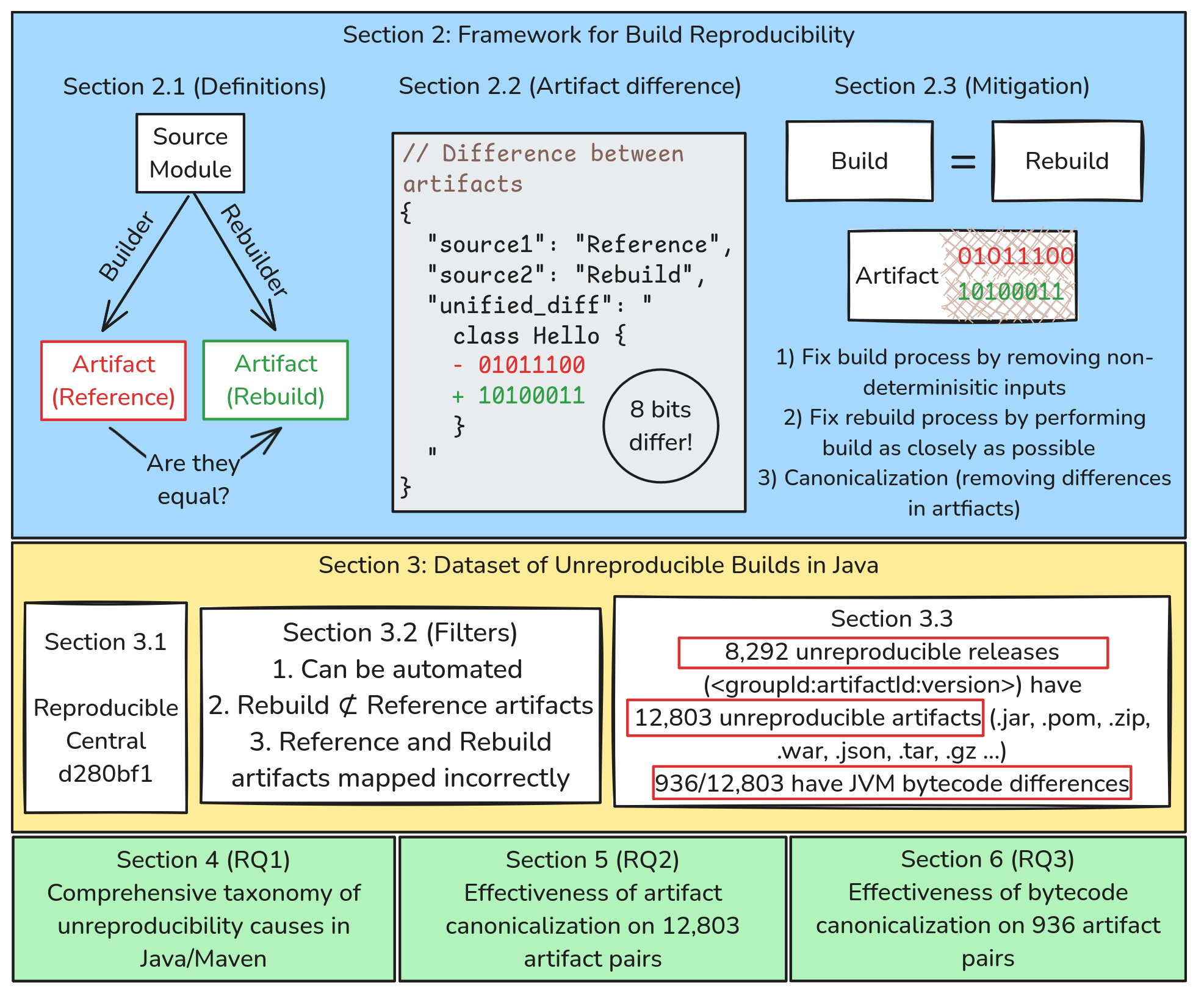}
  \caption{The problem space of unreproducible builds in Java, and our novel mitigation strategies.}
  \label{fig:workflow}
\end{figure*}

\section{Introduction}

The growing complexity of software supply chains~\cite{soto-valero_multibillion_2022,cox_surviving_2019}, coupled with the increasing frequency of supply chain attacks~\footnote{\url{https://www.sonatype.com/hubfs/SSCR-2024/SSCR_2024-FINAL-10-10-24.pdf}},  raises concerns about software integrity.
\newc{In such a fragmented ecosystem, relying solely on assumptions about  the identity of the distributor~\cite{thompson_reflections_1984} is no longer sufficient --- what is needed is verifiable evidence that the software one installs corresponds exactly to its declared source.} 
This challenge is especially acute in open source environments, where binaries are often distributed separately from their source code~\cite{davies_software_2011}, making it difficult for users to independently validate what they are running.
As a result, the focus is shifting toward techniques that can guarantee that what is built matches what was intended, regardless of who performs the build.

This technique is formally called ``Reproducible Build''~\cite{williams_research_2025,cox_fifty_2025}: builds are considered ``reproducible'' if and only if the build process is deterministic, where the same binary can be computed from the same source code by independent parties~\cite{lamb_reproducible_2022}.
It helps prevent attacks on the build process, where an attacker modifies it to insert backdoors or malicious code into the software application to be built~\cite{perry_deterministic_2013}.
Any backdoor or malicious code would be detected by the verifier as the built artifact  would not match the original artifact.
\newc{This approach has already seen adoption in security and privacy critical environments, such as Bitcoin and Tor Browser \cite{levsen_tale_2025}.
Ensuring that clients for both of them are free of any backdoors injected by the build system.}

% the problems of reproducible builds in java
% also essential for java
% citations to paper saying it's a problem
% concrete example on the internet of people who suffer from this
Despite the promises of Reproducible Builds, ensuring and verifying reproducibility at scale remains technically challenging due to the multitude of spurious differences in build outputs (e.g., timestamps, file ordering).
Those differences make binaries appear different even when built from the same source, with an untampered build pipeline~\cite{fourne_its_2023}.
%The US government mandates reproducibility for the software used.
%If a software is not reproducible, an explanation is required~\footnote{\url{https://www.nsa.gov/Press-Room/News-Highlights/Article/Article/3146465/nsa-cisa-odni-release-software-supply-chain-guidance-for-developers/}}.
\newc{Spurious differences hinder reproducibility and create a new attack surface.
Several high-profile supply chain attacks have exploited unreproducible builds.
For instance, in 2024 it was discovered that the official tarballs of XZ Utils, compression library used in almost all Linux distributions, contained a backdoor that enabled remote code execution, even though the backdoor was absent in the project's Git repository~\cite{przymus_wolves_2025}.
Similarly, the Ultralytics Python package, a widely used library for AI models, was compromised when attackers injected malicious code into the build pipeline, resulting in a PyPI release with a backdoor~\cite{larson_supplychain_2024}.
A related incident targeted the Solana ecosystem as well~\cite{claburn_solana_2024}.}
\newc{These incidents share a common vector: the artifacts distributed through registries or release tarballs differed from what could be obtained by building directly from the publicly available source code.
If the build processes for these projects had been reproducible, users or third party rebuilders could have verified that the distributed binaries and packages matched the source.
Any discrepancies would have immediately signaled tampering, thereby preventing such attacks~\cite{malka_how_2025}.}

% our contribution in this paper, our goals, high level goals
In this work, we contribute to solving the problem of achieving build reproducibility in the context of enterprise software in Java.
Recall that Java has consistently been one of the top 5 programming languages in the world for the past 5 years~\footnote{\url{https://innovationgraph.github.com/global-metrics/programming-languages}}, underscoring its widespread use and relevance.
Its importance in the context of finance~\cite{soto-valero_multibillion_2022, morales_java_2021}, government services~\cite{robinson_trends_2017}, and military applications~\footnote{\url{http://pdf.cloud.opensystemsmedia.com/vita-technologies.com/Aonix.Jun07.pdf}}~\footnote{\url{https://tsri.com/news-blog/press/u-s-department-of-defense-mainframe-cobol-to-java-on-aws}} highlights the need for reproducible builds in Java.
We aim at building the most comprehensive  taxonomy of unreproducible builds in Java, and the corresponding mitigation strategies.
Furthermore, we perform a deep study how canonicalization --- removal of non-deterministic differences --- is a good solution for mitigating unreproducibility.

% one paragraph of the overall methodology of the paper, refer to Figure 1
Our approach for analyzing unreproducible builds in Java is shown in \autoref{fig:workflow}.
We first propose an original framework for build reproducibility where we clearly define the roles of the builder and rebuilder and how both of them contribute to reproducible build verification.
Next, we build a dataset of unreproducible builds in Java by leveraging Reproducible Central~\cite{boutemy_jvmreporebuild_2024}, the leading project for rebuilding and verifying Maven applications.
We systematically analyze the dataset to identify and classify the causes of unreproducibility into an original taxonomy of unreproducibility.
Finally, we leverage the dataset to evaluate the effectiveness of bytecode and artifact canonicalization.
Bytecode canonicalization focuses on internal representation of program logic and eliminates compiler introduced variations.
While artifact canonicalization eliminates spurious variations in metadata that can influence how the artifact is interpreted or processed by downstream build tools.
Together, these two levels of canonicalization address both semantic and non-semantic sources of non-determinism, significantly improving overall build reproducibility.

% main findings
% taxonomy
\newc{Our study reveals 6 reasons for unreproducibility in Java, each with the underlying root causes and the best practice mitigations.}
For example, changes in the bytecode in Java artifacts.
% canonicalization
Our large-scale experiments empirically measure the effectiveness of two state-of-the-art canonicalization techniques.
\newc{First, we show that bytecode canonicalization with \jNorm~\cite{schott_java_2024} can successfully canonicalize \jnormsuccess out of \unreproducibleartifactswithjvmbytecode (29.38\%) unreproducible artifacts with JVM bytecode.}
We study the reasons why \jNorm fails to canonicalize the remaining artifacts, which are promising areas of future work.
\newc{Next, we evaluate artifact canonicalization using Google's \ossrebuild and find that it can successfully canonicalize 9.41\% (1,205 / \unreproducibleartifacts) of unreproducible artifacts.}
\newc{By implementing new canonicalization patterns informed by our taxonomy, our new tool \chainsrebuild increases the verification success rate to 26.60\% (\chainsrebuildsuccess / \unreproducibleartifacts)}.
Our paper is the first to demonstrate, at a large scale, that canonicalization is a promising mitigation strategy for fixing build unreproducibility in the Java ecosystem.

% novelty claim paragraph
There has been prior work on reproducible builds in Java focusing on bytecode canonicalization~\cite{dietrich_levels_2024} and fixing build process~\cite{xiong_build_2022}.
However, we are the first to analyze Maven projects hosted on Maven Central at scale and to propose a comprehensive taxonomy of unreproducibility causes and their mitigations.
In addition to bytecode canonicalization, we also study artifact canonicalization and show that it is a promising solution to mitigate unreproducibility in Java. 

To summarize, our contributions are:

\begin{itemize}
    \item A sound conceptual framework of build reproducibility, founded on the novel roles of the  ``builder'' and ``rebuilder'' and on the powerful concept of artifact canonicalization. 
    \item A novel, comprehensive taxonomy of 6 unreproducibility causes and their mitigation in Java, helping practitioners to improve build reproducibility in their projects. 
    \item \newc{A large-scale experiment on the effectiveness of canonicalization, showing that it is a promising solution to mitigate unreproducibility in Java, with artifact canonicalization being more effective at scale than bytecode canonicalization.}
    \item A publicly available, curated dataset of \unreproduciblereleases unreproducible Maven releases for future research.
\end{itemize}

\section{Framework for Build Reproducibility}
\label{sec:theory}

\subsection{Definitions}

Developers often release reusable software in package registries, making them widely available for integration into third-party software projects.
\newc{For our work on build reproducibility, we need to introduce key terminology. We work in the context of the Java programming language and its Maven build system. In the following, we illustrate the key concepts with the example of \href{https://github.com/google/guava}{Apache Guava}.}

\textbf{Package Registry}: A package registry is a centralized repository of software packages that can be reused by other applications.
In this paper Maven Central, hosted by a company called Sonatype, is the considered package registry.

\textbf{Module}: A module is a uniquely addressable component in a software system. In Maven, a module is referenced by a combination of Group Identifier, Artifact Identifier, and Version.
Group ID is a unique identifier for the organization, e.g., ``org.apache'' ~\footnote{\url{https://maven.apache.org/guides/mini/guide-reproducible-builds.html}}.
Artifact ID is the unique name of the module and is unique within the group ID.
For example, \texttt{\path{com.google.guava:guava:32.0.0-jre}} is a Maven module where \texttt{\path{com.google.guava}} is the group ID, \texttt{\path{guava}} is the artifact ID, and \texttt{\path{32.0.0-jre}} is the version.
There are two types of modules: a \textbf{source module} is a module that is hosted on a version control system (e.g., Github), not necessarily published on a package registry; a \textbf{released module} is one published on a package registry, Maven Central in this paper.
\newc{Both released and source modules can be identified by the combination of group ID, artifact ID, and version. In the latter case, we assume the presence of a mapping from version to Git tag or Git commit.}

\textbf{Project}: A Maven project is a collection of source modules in the same source repository.
It contains build configuration for all the modules in a so-called `POM file'.
\newc{For example, the Maven project \texttt{\path{guava}} contains 6 source modules --- \texttt{\path{guava-parent}}, \texttt{\path{guava-tests}}, \texttt{\path{guava-testlib}}, \texttt{\path{guava-gwt}}, \texttt{\path{guava-bom}}, and \texttt{\path{guava}}.}

\textbf{Artifact}: An artifact is any file that is part of the released module.
In Maven, there are mostly a JAR or a ZIP archive.
For example, in released \texttt{\path{com.google.guava:guava:32.0.0-jre}}, \texttt{\path{guava-sources.jar}} and \texttt{\path{guava.jar}} are JAR file in the released module.
In this paper, we have two types of artifacts:
The \textbf{reference artifact} is the artifact of the released module that is available on the package registry;
The \textbf{rebuild artifact} is the artifact obtained by rebuilding the source module.

\textbf{Builder}: The builder is the entity that publishes a released module on a package registry, from the source module.
A builder is often an automated script implemented in a CI/CD pipeline, like GitHub Actions.
For example, the released module \texttt{\path{com.google.guava:guava:32.0.0-jre}} has been deployed on Maven Central by the Google builder.
%at this URL \url{https://repo.maven.apache.org/maven2/com/google/guava/guava/32.0.0-jre/}.
% The URL stores all the artifacts belonging to the release.

\textbf{Rebuilder}: The rebuilder is the entity that verifies the build reproducibility of the released module, with information available from the package registry.
In other words, the rebuilder 1) builds the projects again 2) verifies the reproducibility of all the rebuild artifacts against the artifacts in the released module.
For example, the Reproducible Central project~\cite{boutemy_jvmreporebuild_2024} is a rebuilder that verifies the reproducibility of the released modules of 700+ open-source Java libraries.

\textbf{Canonicalization}
Canonicalization is the process of converting different representations of the same data into a single, standard representation.
For example, in the domain of XML parsing~\cite{boyer_canonical_2008}, canonicalizing means encoding the XML document in a single charset (UTF-8), adding default attributes, and many more changes as listed in the specification~\cite{boyer_canonical_2008}.

Canonicalization is useful in the context of build reproducibility as it allows to convert the output artifacts of the build process into a standard representation that is independent of the machine and environment where the build has been performed.
Assume that the output artifact contains details of the operating system or the user who built the artifact. These details are not relevant to the functionality of the software and can be removed as part of canonicalization or the artifact.

\subsection{Build Reproducibility}

Build Reproducibility is a property of a software build process where the output artifact is bit-by-bit identical when built again, given a fixed version of source code and build dependencies, regardless of the environment~\cite{lamb_reproducible_2022}.
This is shown in the equation below,  where source module is built twice by builder and rebuilder to produce reference and rebuild artifacts, respectively.
The build is reproducible if the reference and rebuild artifacts are identical.

Reproducible builds help prevent software supply chain attacks at the build time by ensuring that the software is not tampered with during the build process.
It also allows users to verify that the executable provided to them is free of backdoors as they can build the software themselves and compare the output with the provided executable.

\begin{align}
  \label{eq:reproducible-build}
    Reference Artifact \leftarrow Build(Source Module)  \\
   Rebuild Artifact \leftarrow Rebuild(Source Module) \\
  ReferenceArtifact = RebuildArtifact
\end{align}

There are two other terms in the literature that are used interchangeably with build reproducibility --- ``verifiable builds'' and ``accountable builds''~\cite{pohl_sok_2025}.
Builds are verifiable if the build process can be modified in order to make the output artifact bit-by-bit identical.
Accountable builds are the same as verifiable builds, but instead of removing the differences, these differences are explained and documented so that unreproducibility can be understood upon inspection.
In our paper, we stick with the term ``build reproducibility'' as the other two terms are extensions of build reproducibility with extra steps.

\subsection{Mitigation}
\label{sec:mitigation}

\newc{Our background study on fixing reproducibility (see ~\autoref{sec:fixing-unreproducibility}), shows that any cause of unreproducibility can be mitigated in three ways --- fixing the build process, fixing the rebuild process, or canonicalizing the output of the rebuild process.}

\textbf{Fixing the build process} and \textbf{Fixing the rebuild process} involve modifying inputs such as source code, dependencies, or build tools (e.g., build scripts and configurations) to the build or rebuild workflows in order to eliminate sources of non-determinism.
The builder ensures that the build process is deterministic and easy to reproduce for rebuilder.
The rebuilder tries to perform the build process as closely as possible to the original build process.
However, some build tools generate non-deterministic information or there is a difference in the environment which makes these mitigation techniques not always feasible neither practical.

\textbf{Canonicalizing the output of the rebuild process} as defined above means removing non-deterministic information from output artifacts.
Some non-deterministic information is spurious, but it is sometimes perfectly legitimate such as a cryptographic signature with one-time key \cite{decarnedecarnavalet_challenges_2014}.
Canonicalization is generally not done by the builder for this reason and it is more the responsibility of the rebuilder.
It requires careful removal of all non-deterministic information from the output artifact.
If done incorrectly, it may inadvertently remove or obscure meaningful differences in the output artifacts.
This can lead to false positives, where builds that are genuinely unreproducible are incorrectly reported as reproducible.
Such errors would undermine the reliability of reproducibility verification and can mask underlying issues that need to be addressed.
Therefore, it is crucial to ensure that canonicalization is performed with precision, preserving all relevant differences while eliminating only the spurious, non-deterministic variations.
Canonicalization can address the unreproducibility issues of past releases by removing non-deterministic information from its output artifacts.
It can also help as last resort to fix unreproducibility issues when build process and rebuild process are embedding non-deterministic information which is not in control (like cryptographic signatures).
Moreover, techniques related to fixing build process have also been studied in prior literature~\cite{randrianaina_options_2024,ren_automated_2022,mukherjee_fixing_2021b,xiong_build_2022,keshani_aroma_2024,macho_nature_2021}.
Although we suggest these techniques as mitigation strategies, we focus deeply on the third mitigation strategy (canonicalization) due to the fact that it is not well evaluated in the literature.

\section{Dataset of Unreproducible Builds}
\label{sec:dataset}

\begin{table}[h]
  \centering
  \small
  \begin{tabular}{|l|c|c|}
    \hline
    \multirow{2}{*}{\textbf{Dataset Stage}} & \textbf{Projects} & \textbf{Releases} \\
                 & \textbf{Count} & \textbf{Count} \\
    \hline
    Initial Reproducible Central (\href{https://github.com/jvm-repo-rebuild/reproducible-central/tree/d280bf1555e2ec6f172b678017b7f00370cb7a00}{d280bf1}) & 706 & 4,956 \\
    After filters in \autoref{sec:java-projects-filter} & 684 & 4,030 \\
    After filters in \autoref{sec:released-modules-filter} & 271 & 1,144 \\
    \hline
  \end{tabular}
  \caption{\newc{Evolution of dataset size through filtering stages, showing the reduction from initial Reproducible Central data to final unreproducible releases.}}
  \label{tab:dataset-evolution}
\end{table}

\subsection{The Reproducible Central Project}

\newc{The Reproducible Central project~\cite{boutemy_jvmreporebuild_2024} is a rebuild infrastructure to verify whether Java projects are reproducible.}
It specifies list of steps to build a Java project from source in a \texttt{\path{buildspec}} file~\footnote{\url{https://github.com/jvm-repo-rebuild/reproducible-central/blob/master/doc/BUILDSPEC.md}}.
Then, the Reproducible Central script takes this \texttt{\path{buildspec}} as input to rebuild a project.
It produces \texttt{\path{buildinfo}} and \texttt{\path{buildcompare}} files;
\texttt{\path{buildinfo}} records the output of the build in terms of the artifact names, sizes, and checksums~\cite{builds_jvm_2024};
\texttt{\path{buildcompare}} reports which artifacts are reproducible and which are unreproducible.

We use the Reproducible Central project dataset as a starting point to study the causes of unreproducible builds in Java.
We take a snapshot of the Reproducible Central dataset on 8th October, 2024 whose commit is \href{https://github.com/jvm-repo-rebuild/reproducible-central/tree/d280bf1555e2ec6f172b678017b7f00370cb7a00}{d280bf1}.
At this date, Reproducible Central contains 706 projects, their corresponding build scripts, and a history of 4,956 releases.
It is the largest dataset attempting rebuilds of Java projects and is also maintained by maintainers of Apache Maven.

\newc{Our goal is to curate a dataset of unreproducible released Java modules in order to study the reasons why unreproducible builds occur.}
% Reproducibility of each released module is more important than the entire project as released modules are either used directly or declared as dependency and projects are not. If reproducibility check is mapped to each released module, the consumer can know if they are executing reproducible software or the dependency they declare is reproducible or not.
\newc{In order to curate the dataset of unreproducible released modules, we first apply one filter before running the rebuilds and two filter afterwards.}
% Then we convert all the unreproducible projects into unreproducible released modules.

\subsection{Filtering}

\newc{\autoref{tab:dataset-evolution} shows the evolution of the dataset size through filtering stages.
We start with 706 Java projects and 4,956 releases in the Reproducible Central dataset.
It contains 4,851, 103, and 2 releases built with Maven\footnote{\url{https://maven.apache.org/}}, Gradle\footnote{\url{https://gradle.org/}}, and SBT \footnote{\url{https://www.scala-sbt.org/}}, respectively.
Next, we apply two filters to the dataset which reduces the dataset to 684 Java projects and 4,030 releases.
Finally, we only keep the unreproducible releases and projects which results in 271 Java projects and 1,144 releases.}

\newc{Next, we account for the fact that the 271 Java projects are made of multiple modules. Consequently, we further split the 271 Java projects and their 1,144 releases into \unreproduciblereleases unreproducible module releases as we account for unreproducibility per <GroupId, ArtifactId, Version> coordinates of Java modules rather than the entire source project.}

\subsubsection{\newc{Filter Java Projects}}
\label{sec:java-projects-filter}
\newc{We filter out Java projects that require manual intervention to build. This is to ensure that we can run the entire rebuild process automatically.}
% precise version of JDK is removed because it seemed easy to automate.
This is indicated in \texttt{\path{buildspec}} file either by the presence of build tool or when the rebuild command is prefixed with `SHELL'.
The rebuild script would spawn an interactive terminal when any of the above conditions are met.
%\href{https://github.com/jvm-repo-rebuild/reproducible-central/blob/d280bf1555e2ec6f172b678017b7f00370cb7a00/content/com/io7m/hibiscus/com.io7m.hibiscus-0.0.3.buildspec}{com.io7m.hibiscus-0.0.3.buildspec} show that build requires \texttt{\path{mvn-3.9.3}} build tool and \texttt{\path{17.0.6}} version of JDK. \href{https://github.com/jvm-repo-rebuild/reproducible-central/blob/d280bf1555e2ec6f172b678017b7f00370cb7a00/content/org/apache/pulsar/pulsar-3.0.2.buildspec}{pulsar-3.0.2.buildspec} shows how the `SHELL' keyword is prefixed to the rebuild command.
% this excludes 106 maven projects / 4963
We exclude projects that do not have a POM file at the root directory as it also requires manual intervention to get the path to rebuild artifacts in build directory.
%We automated this process later in a pull request but do not include this fix as it is not part of the snapshot~\footnote{\url{https://github.com/jvm-repo-rebuild/reproducible-central/pull/1817}}.
\newc{Once all rebuilds are done, we apply the second filter that excludes projects where the unreproducible artifacts are not a subset of all the artifacts released on Maven Central.}
This can happen when not all artifacts produced by the build process are released on Maven Central.
For example, \texttt{\path{content/org.infinispan:protostream-aggregator:5.0.7.Final}} produces an artifact \texttt{\path{rotostream-integrationtests-5.0.7.Final-test-sources.jar}} which is not released on Maven Central.
% \newc{After applying the two filters, we end up with 3,927 releases over 665 Maven projects.}
\newc{After applying the two filters, we end up with 4,030 releases over 684 Java projects.}

\subsubsection{Filter Released Modules}
\label{sec:released-modules-filter}
Next, we need to identify the released modules to compare against.
\newc{In order to map each unreproducible project to corresponding released modules, we first identify the Java submodules in each project.}
\texttt{\path{org.apache.aries.cdi:org.apache.aries.cdi.executable:1.1.5}} is an example of a source module that does not have a corresponding URL on Maven Central.
We parse the POM file located at the root directory of the project as it contains the required information about the modules.
\newc{For example, jpmml-python is a Maven project hosted on GitHub~\footnote{\url{https://github.com/jpmml/jpmml-python/tree/1.2.2}} which has three Maven modules with version \texttt{\path{1.2.2}} --- \texttt{\path{org.jpmml:jpmml-python}}, \texttt{\path{org.jpmml:pmml-python:1.2.2}}, and \texttt{\path{org.jpmml:pmml-python-testing}}}.
Thus, jpmml-python Maven project on Reproducible Central corresponds to three Maven modules in our dataset.
Two out of three released modules in this project contain unreproducible artifacts.
\texttt{\path{org.jpmml:pmml-python-testing:1.2.2}} has \texttt{\path{pmml-python-testing-1.2.2-sources.jar}} and \texttt{\path{org.jpmml:pmml-python:1.2.2}} has \texttt{\path{pmml-python-1.2.2-sources.jar}}.
Finally, there are four 4 cases where unreproducible reference and rebuild artifact are not mapped correctly.
Maven module \texttt{\path{org.apache.ratis:ratis-assembly:3.1.0}} and three more of its versions have an artifact with \texttt{\path{.pom}} extension.
The rebuild script maps this artifact to the Jar file produced during rebuild.
\newc{The third and final filter excludes these four cases.}
This is a bug in the Reproducible Central rebuild script and we have reported this to the author~\footnote{\url{https://github.com/jvm-repo-rebuild/reproducible-central/issues/2164}}.

\newc{At the end of the pipeline, we end up with \unreproduciblereleases released Java modules which have a total of \totalartifactsinallunreproduciblereleases artifacts from 4,030 releases of 684 projects.}
\newc{Finally, we end up with 271 Java projects and 1,144 releases which are unreproducible and they correspond to \unreproduciblereleases unreproducible released modules that have \unreproducibleartifacts/\totalartifactsinallunreproduciblereleases (35.8\%) unreproducible artifacts.}

\subsection{Description of the Dataset}

The dataset contains groupId, artifactId, version, and the URL of all \unreproduciblereleases unreproducible released modules on Maven Central.
It also contains the reference and rebuild artifact pair of all \unreproducibleartifacts unreproducible artifacts.
It is available at \url{https://github.com/chains-project/reproducible-central/}.
To our knowledge, this is the largest ever curated dataset of unreproducible artifacts in the context of Java.
Based on the dataset, we can answer the following research questions:
\begin{itemize}
    \item RQ1 (Causes): \rqtaxonomy
    \item RQ2 (Artifact Canonicalization): \rqossrebuild
    \item RQ3 (Bytecode canonicalization): \rqjnorm
\end{itemize}

\section{RQ1: Taxonomy of Unreproducibility Causes}
\label{sec:taxonomy}

\subsection{Objective}
The goal of this section is to provide a comprehensive taxonomy of unreproducibility causes in Java.
We first report the cause based on our analysis and then propose a mitigation strategy for each cause.

\subsection{Methodology}

We have a dataset of \unreproducibleartifacts unreproducible artifacts and for each of them we have the reference artifact from Maven Central and its rebuilt version from our own rebuild.
We run diffoscope 285 on each pair of reference and rebuild artifacts to get the content diff.
\diffoscope is a state-of-the-art tool for computing the difference between various types of files, archives, and directories~\footnote{\url{https://diffoscope.org/}}.
This gives us a total of \unreproducibleartifacts diffoscope files.

Given the scale of the dataset, we first analyze a random sample of the diffoscope files to understand the reasons of unreproducibility.
There are two attributes in the diffoscope file that are of interest to us --- \texttt{\path{source1}} and \texttt{\path{unified_diff}}, both of them indicating a reproducibility problem.
The \texttt{\path{source1}} attribute either contains the name of the file or the command that is run to get the textual output of the file.
For example, if \texttt{\path{source1}} includes \texttt{\path{cyclonedx.json}}, we can infer that the difference is in the Software Bill of Materials (SBOM).
Another example is if \texttt{\path{source1}} is \texttt{javap -verbose -constants -s -l -private \{\}}, it means that the Java class file is being compared and the reason of unreproducibility is due to differences in the bytecode.
The \texttt{\path{unified_diff}} attribute contains the diff of the file in the unified format, for textual files.
\newc{Next, we notice that the changes are present in the most deeply nested \texttt{\path{unified_diff}} attribute which represents the textual difference in the file extracted from the archive.}
\newc{This is because Diffoscope recursively extracts files from the archive and shows the difference in the file.}
\newc{The difference in archive indicates that there is a difference in one of its file, and the textual difference in that particular file can unveil the cause of unreproducibility.}
\newc{Thus, we reduce the search space to the most deeply nested \texttt{\path{unified_diff}} attribute.
This also helps us to easily read the difference in the file.}

\newc{We then write regular expressions to cluster the causes of unreproducibility based on file names and tool names reported in the \texttt{\path{source1}} attribute and specific pattern in the \texttt{\path{unified_diff}} attribute.}
\newc{For example, if regex}
\begin{verbatim}
  ^[+-].*\d{4}-\d{2}-\d{2}T\d{2}:\d{2}:\d{2}\.\d{3}Z$
\end{verbatim}
\newc{is matched, we infer that the reason of unreproducibility is due to timestamps in the file and then we manually inspect the difference in the file to validate the root cause of the unreproducibility.}
\newc{The manual inspection also helps us to identify more reasons for unreproducibility.
For example, while looking through differences in timestamps, we also observe differences about version control system information.
This iterative process through manual analysis continuously refines the taxonomy.}

\newc{It is important to note that regular expressions are not directly mapped to the reasons we propose later, because this would have too many false positive.
We rather rely on manual analysis to categorize differences.}
\newc{Thus, we do not count the causes of unreproducibility but build a comprehensive taxonomy about Java unreproducibility.}

\subsection{Results}

\begin{table*}[h!]
  \centering
  \small
  \begin{tabular}{|l|l|p{3.5cm}|c|l|}
  \hline
  \textbf{Reason for Unreproducibility} & \textbf{Root Cause of Unreproducibility} & \textbf{Fine-grained Cause} & \textbf{Novelty} & \textbf{Main Mitigation} \\ \hline
  
  \multirow{12}{*}{Build Manifest} 
    & \multirow{3}{*}{Environment} 
      & Built-By &  & \multirow{4}{*}{Canonicalization by rebuilder} \\ \cline{3-4}
    & & Signed JARs & \cmark & \\ \cline{3-4}
    & & Os-Version & \cmark & \\ \cline{3-4}
    & & Eclipse Properties & \cmark & \\ \cline{2-5}
  
    & \multirow{5}{*}{Rebuild Process} 
      & Implementation-Build-Java-Vendor & \cmark & \multirow{5}{*}{Fix rebuild process} \\ \cline{3-4}
    & & Created-By & \cmark & \\ \cline{3-4}
    & & Originally-Created-By & \cmark & \\ \cline{3-4}
    & & SCM-Revision & \cmark & \\ \cline{3-4}
    & & Build-Jdk & & \\ \cline{2-5}
  
    & \multirow{2}{*}{Dynamic Properties}
      & SCM-Git-Branch & \cmark & \multirow{2}{*}{Fix build process} \\ \cline{3-4}
    & & Bnd-LastModified & \cmark & \\ \cline{2-5}
    & Inconsistent Build Configuration
      & Order of properties and their values & & Canonicalization by rebuilder \\ \hline
  
    \multirow{8}{*}{Software Bill of Materials}
    & Java Vendor & Removal of hash algorithms & \cmark & Fix rebuild process \\ \cline{2-5}
    & \multirow{3}{*}{Inconsistent Build Configuration} 
      & Addition of components & \cmark & \multirow{3}{*}{Fix build process} \\ \cline{3-4}
    & & Removal of components & \cmark & \\ \cline{3-4}
    & & Modification of components & \cmark & \\ \cline{2-5}
    & \multirow{2}{*}{Dynamic Properties}
      & Timestamp & \cmark & \multirow{2}{*}{Canonicalization by rebuilder} \\ \cline{3-4}
    & & SerialNumber & \cmark & \\ \cline{2-5}
    & \multirow{3}{*}{External Metadata}
      & License & \cmark & \multirow{3}{*}{Canonicalization by rebuilder} \\ \cline{3-4}
    & & Description of components & \cmark & \\ \cline{3-4}
    & & ExternalReferences & \cmark & \\ \hline
  
    \multirow{6}{*}{Filesystem}
    & \multirow{4}{*}{Environment}
      & Permissions & \cmark & \multirow{4}{*}{Canonicalization by rebuilder} \\ \cline{3-4}
    & & Ownership & \cmark & \\ \cline{3-4}
    & & Size & \cmark & \\ \cline{3-4}
    & & Absolute Paths & \cmark & \\ \cline{2-5}
    & \multirow{2}{*}{Inconsistent Build Configuration}
      & Type & \cmark & \multirow{2}{*}{Fix build process} \\ \cline{3-4}
    & & Files removed or added & \cmark & \\ \hline
  
    \multirow{7}{*}{JVM Bytecode}
    & \multirow{3}{*}{JDK Version}
      & Debug information & \cmark & \multirow{3}{*}{Fix build process} \\ \cline{3-4}
    & & Optimization and de-optimization & & \\ \cline{3-4}
    & & Refactoring & \cmark & \\ \cline{2-5}
    & \multirow{1}{*}{Embedded Metadata}
      & Absolute file paths, timestamps, Java/project version, Git properties, usernames & \cmark & Fix build process \\ \cline{2-5}
    & \multirow{2}{*}{Generated Code}
      & Java Compiler &  & Canonicalization by rebuilder \\ \cline{3-5}
    & & Maven Plugins & \cmark & Fix rebuild process \\ \hline

    \multirow{6}{*}{Versioning Properties}
    & \multirow{6}{*}{Source Repository State}
      & Number of commits &  & \multirow{6}{*}{Canonicalization by rebuilder} \\ \cline{3-4}
    & & Number of Git tags & & \\ \cline{3-4}
    & & Branch name & & \\ \cline{3-4}
    & & Local branch name & & \\ \cline{3-4}
    & & Timezone of commit & & \\ \cline{3-4}
    & & Remote URL & & \\ \hline
  
    \multirow{6}{*}{Timestamps}
  & \multirow{6}{*}{Build-time Variability}
    & Documentation & \cmark & \multirow{6}{*}{Fix build process} \\ \cline{3-4}
  & & Shell scripts & \cmark & \\ \cline{3-4}
  & & Executable binaries & \cmark & \\ \cline{3-4}
  & & JVM bytecode & \cmark & \\ \cline{3-4}
  & & Build manifest &  & \\ \cline{3-4}
  & & File metadata & \cmark & \\ \hline
  
  \end{tabular}
  \caption{\newc{The taxonomy of unreproducibility causes based on the analysis of \unreproducibleartifacts unreproducible artifacts. We are the first to study the most appropriate mitigation strategy.}}
  \label{tab:taxonomy}
\end{table*}

In this section, we present a comprehensive taxonomy of unreproducibility in Java, based on our large-scale, systematic protocol presented in~\autoref{fig:workflow}.
\autoref{tab:taxonomy} summarizes the taxonomy.
\newc{The first column lists the reason for unreproducibility.}
\newc{The second column lists the root cause that triggered the difference.}
\newc{The third column further breaks down the root cause into fine-grained causes.}
\newc{The fourth column indicates if the root cause is novel and has not been reported for Java artifacts in the literature before, including recent work by Xiong et al.  \cite{xiong_build_2022}.}
\newc{Finally, the last column suggests the main strategy to fix the unreproducibility, one of the three presented in \autoref{sec:mitigation}.}

\newc{For example, consider the row where `JVM Bytecode' is the reason and `Generated Code' is the root cause.
We break it down into two further causes - `Java Compiler' and `Maven Plugins'.
A build difference due to generated JVM bytecode by Maven plugins has not been reported in the literature before.
However, a build difference due to Java compiler has been reported by Xiong et. al~\cite{xiong_build_2022}.}

\newc{For differences due to Java compiler, we can use \jNorm which canonicalizes the differences.
Differences due to Maven plugins can be mitigated by fixing the rebuild process.
%As an example, we observe custom build configuration have not been included in the rebuild process. The rebuilder can automatically infer such configurations from the build manifest and run the correct build process.
}

\subsubsection{\textbf{Unreproducible Build Manifests}}

We consider two manifest files in our analysis --- \texttt{\path{MANIFEST.MF}} and \texttt{\path{pom.properties}}.
The MANIFEST.MF file is a mandatory metadata file in a JAR file.
It potentially contains information on the entry point of the JAR file, signatures of the classes, and details of the version control system~\footnote{\url{https://docs.oracle.com/javase/tutorial/deployment/jar/manifestindex.html}}.
\texttt{\path{pom.properties}} is an automatically generated file by Maven that contains the group ID, artifact ID, and version of the Maven module. 

The first set of differences in manifests is due to a change in environment.
For example, the \texttt{\path{Built-By}} attribute in the \texttt{\path{MANIFEST.MF}} file is sometimes updated to the username of the user who built the JAR file.
\textbf{Mitigation:} This non-deterministic information should be stripped out from the JAR file by rebuilder before comparison.
See \cref{appendix:mf-built-by}, \cref{appendix:mf-signed-jars}, \cref{appendix:mf-os-version}, and \cref{appendix:pom-properties-eclipse} for more attributes that follow the same problem and solution.

The next set of differences in manifests is due to differences in the script that configures the rebuild.
For example, the \texttt{\path{Build-Jdk}} attribute depends upon the version of JDK used to build the JAR file.
\textbf{Mitigation:} All environment specifications must be exactly specified in the the build script by rebuilder.
\newc{Otherwise, the difference is not only in the attribute but it impacts other components of the release too.}
\newc{For example, the \texttt{Build-Jdk} attribute in the \texttt{\path{MANIFEST.MF}} file is set to the version of JDK used to build the JAR file and the rebuilder should use the same version of JDK to rebuild the JAR file.
As we will see later, a different JDK versions can lead to differences in the JVM bytecode and hence the artifact becomes unreproducible.}
See \cref{appendix:mf-implementation-build-java-vendor}, \cref{appendix:mf-build-jdk}, \cref{appendix:mf-created-by}, \cref{appendix:mf-originally-created-by}, and \cref{appendix:mf-scm-revision} for concrete examples.

Finally, there are reasons of differences in the manifests \texttt{\path{MANIFEST.MF}} and \texttt{\path{pom.properties}} that are due to non-deterministic behavior in the build script of the builder.
For example, build manifests \texttt{\path{pom.properties}} and \texttt{\path{MANIFEST.MF}} can differ in order of the attributes if they are not sorted.
\newc{\textbf{Mitigation:} The builder should depend  on plugins that generate sorted data~\footnote{\url{https://github.com/apache/maven-archiver/commit/763a940540eefad74f9ba73cb5eed288dc4e639d}}, should not embed non-deterministic data (e.g., branch names, timestamps) in the artifact, and stay up to date with the latest version of plugins.
For example, only the latest version of Maven Archiver is able to maintain a deterministic file  ordering.}
\cref{appendix:mf-scm-git-branch}, \cref{appendix:mf-bnd-lastmodified}, \cref{appendix:mf-order}, \cref{appendix:pom-properties-order}, and \cref{appendix:pom-properties-timestamp} describe these reasons and solutions in detail.

\subsubsection{\textbf{Unreproducible Software Bill of Materials}}

Software bill of material (SBOM) is a formal, machine-readable inventory of software components, metadata about those components, and the hierarchical relationships between them~\footnote{\url{https://www.ntia.gov/sites/default/files/publications/sbom_at_a_glance_apr2021_0.pdf}}.
Its goal is to enable transparency of the software supply chain of an application, for its consumers.
In the case of Java, the best practice is to push SBOMs as part of the Maven release~\cite{gamage_software_2025}.
Our experiment finds that this poses reproduciblity challenges.
 
The first set of changes are due to incorrect Java vendor used by the rebuilder.
For example, old JDK versions do not have newer hash algorithms like SHA-3.
However, vendors of some JDKs backport these algorithms and if builder uses a JDK with backported algorithms, the resulting SBOM will have hashes computed using more algorithms.
If rebuilder does not use the same JDK, there will be differences in number of hash algorithms used.
\textbf{Mitigation:} The rebuilder should use the same JDK vendor and version as the builder.
\newc{This information may be present\footnote{This is not a standard practice. These properties could either be embedded in MANIFEST.MF or \texttt{module-info.java}, or documented in the README of the source repository. Hence, this information may not always be available and its deduction can be adhoc.} in the \texttt{\path{Build-Jdk}} and \texttt{\path{Implementation-Build-Java-Vendor}} attribute in the \texttt{\path{MANIFEST.MF}} file.}
See \cref{appendix:sbom-hash-algo} for more details.

\newc{The next set of changes is due to a inconsistent build configuration by the builder.}
\newc{For example, the build process may use different ad-hoc scripts or rely on local caches to fetch components.}
\textbf{Mitigation:} \newc{The builder should use a conventional build process, such as only using standard Maven commands if only Maven projects are present in the project.
Caches, like local Maven repository, should also be cleared because it can contain outdated components.
This helps the rebuilder to reproduce the build process exactly without knowing all ad hoc scripting in the build.}
See \cref{appendix:sbom-components-add-remove} and \cref{appendix:sbom-components-modification} for concrete examples.

\newc{The third set of changes is due to volatile identifiers in the SBOM.}
CycloneDX attributes like \texttt{\path{timestamp}} and \texttt{\path{serialNumber}} are non-deterministic and hence they differ in each build.
\newc{\textbf{Mitigation:} These attributes only act as an identifier and hence they can be canonicalized by the rebuilder before comparison.}
\newc{See \cref{appendix:sbom-serial-number} and \cref{appendix:sbom-timestamp} give concrete examples of releases falling into this case.}

\newc{The last set of changes is due to external metadata in the SBOM.
These attributes are fetched from external sources and hence if the external source updates metadata like license, description of component, or external references, the SBOM will be different for the subsequent builds.}
\textbf{Mitigation:} \newc{These three attributes can be canonicalized before comparison. In general, data from mutable external sources should be canonicalized.}
\newc{See \cref{appendix:sbom-external-reference}, \cref{appendix:sbom-description-components}, and \cref{appendix:sbom-license} which lists releases falling into this case.}

\newc{We are the first ones to report that SBOM can be a reason of unreproducible builds.}

\subsubsection{\textbf{Unreproducible Filesystem}}

We observe unreproducible Maven releases due to presence or absence of files, changes in the absolute file paths and their related metadata such as permissions, ownership, size and type.
We classify them into two reasons below.

The first set of changes are due to an influence from the environment.
For example, the permissions of the file can vary depending on the \texttt{\path{umask}} of the environment of rebuilder~\cref{appendix:file-permissions}.
\newc{The ownership and timestamp of the file can also vary depending who and when the artifact is built~\cref{appendix:file-ownership}~\cref{appendix:file-timestamps}.}
\newc{File sizes can vary due to newlines as operating systems handle it differently~\cref{appendix:file-size}.} 
The names of the files can vary if absolute paths are used to refer to the files in artifact~\cref{appendix:file-names}.
These names can be of generated files or they could be embedded in JVM bytecode and configuration files which makes the artifact unreproducible~\cref{appendix:file-embedded}.
Finally, the order of files in Jar file~\cref{appendix:order-files-archive} can vary as the order depends upon the filesystem layout~\footnote{\url{https://reproducible-builds.org/docs/archives/}}.
\textbf{Mitigation:} \newc{The rebuilder should either set a fixed value or strip this information from the metadata of the file.}
\newc{For example, the byte buffer of zip file can be edited to set the username and group of an entry in Jar file to an empty string.}

Finally, there are some causes due to differences in the build process by the builder and rebuilder.
For example, some binaries are not included in the rebuild version as they are not generated by the rebuilder build process.
\textbf{Mitigation:} The rebuilder process should replicate the build process of the builder exactly.
For the above example, the rebuilder should incorporate commands to generate the missing binaries.
\newc{An automatic way to infer the correct process is to look for different Maven profiles and activate them during the rebuild process.}
See \cref{appendix:files-removed-added} and \cref{appendix:file-type} for more concrete examples.

\subsubsection{\textbf{Unreproducible JVM bytecode}}
\label{sec:jvm-bytecode}

We observe \unreproducibleartifactswithjvmbytecode unreproducible Maven releases due to changes in the JVM bytecode when rebuilding.
The reasons are as follows.

The first set of changes are due to the use of different versions of the JDK or JDK flags used in the build and rebuild process.
This differences are due to core compilation changes such as refactoring, ordering changes in the bytecode (specifically methods, fields, static initializers, entries in the constant pool, and array type values in annotation~\cref{appendix:order-jvm}), optimization flags, debug flags.
For example, \texttt{\path{java.nio.ByteBuffer.flip()}} returns a different type in JDK 8 and JDK 11.
For \texttt{\path{io.dropwizard.metrics:metrics-collectd:4.1.20}} rebuilt with JDK 11, all \texttt{\path{flip}} calls return \texttt{\path{java.nio.ByteBuffer}} which is different from the reference version built with JDK 8 where the return type is \texttt{\path{java.nio.Buffer}}.
\footnote{This is reported in the bug tracking system of OpenJDK~\cite{gafter_jdk4774077_2002} which shows that this change is introduced in JDK 9.}
See \cref{appendix:jvm-debug-info}, \cref{appendix:jvm-optimizations}, and \cref{appendix:jvm-refactoring} for more examples.
\textbf{Mitigation:} The rebuilder should use the same JDK version and flags as the builder.
We propose that the builder should always embed the precise vendor and version of JDK used to build the artifact in the MANIFEST file under the attribute \texttt{\path{Build-Jdk}}.
This way, the rebuilder can use the same version and vendor of JDK to rebuild the artifact by reading the information instead of guessing it.
If the builder is using a custom JDK as done in Google Guava~\cite{moore_guava_2023}, this should be made available to the rebuilder for verification.

The next set of changes are due to embedding information related to environment.
We observe that the bytecode has git branch names, user name of the builder, and absolute paths.
For example, \texttt{\path{org.apache.hive:hive-exec:4.0.0-alpha-2}} embeds git properties and timestamps in \texttt{\path{package-info.class}} file using a shell script that adds a Java annotation to \texttt{\path{package-info.java}}.
The Java compiler also embeds the exact Java version used to compile the source code in the bytecode of module-info.
However, this hinders reproducibility even if the patch version of the JDK is different.
\textbf{Mitigation:} The builder should avoid embedding environment specific information in the bytecode.
This information changes with each rebuilder and hence the bytecode will be different making the artifact unreproducible.
See \cref{appendix:embedded-data} for example of a release where it happens.

Finally, the last set of changes are due to build time code generation or modification.
The Java compiler produces synthetic code during build in order to handle access to type members~\cite{rimenti_synthetic_2018}.
The compiler also names lambda functions~\cite{evans_scenes_2020} and anonymous classes~\cite{maverik_why_2016} which are not present in the source code.
In addition to these capabilities, Maven provides plugins to generate or modify code during the build process.
All of this generated code can be different across different builds and hence the build is unreproducible.
For example, we observe a difference in names of lambda function in \texttt{\path{org.apache.helix:zookeeper-api:1.0.3}}.
Another example is \texttt{\path{org.apache.hive:kafka-handler:4.0.0-alpha-2}} which uses \texttt{\path{maven-shade-plugin}} to rename package names from \texttt{\path{kafkaesqueesque}} to \texttt{\path{kafkaesque}} but only when a Maven profile is activated.
\textbf{Mitigation: }The names of the synthetic code can be canonicalized by the rebuilder before comparison.
Bytecode canonicalization~\cite{sharma_sbomexe_2024} is a technique to strip out non-deterministic information from the bytecode.
\newc{Maven can be configured to have multiple build profiles and this changes the artifacts produced.}
\newc{The rebuilder can infer the build profiles required to reproduce the build process by looking at the manifest file (pom.xml) of the artifact and choose the appropriate one which is designed for deployment.}
See \cref{appendix:build-time-generated-code} for more examples.

\subsubsection{\textbf{Unreproducible Versioning Properties}}

Git properties are embedded in Jar files in \texttt{\path{git.json}} or \texttt{\path{git.properties}} files.
\newc{Maven plugin \texttt{\path{git-commit-id-maven-plugin}}~\footnote{\url{https://github.com/git-commit-id/git-commit-id-maven-plugin}} is commonly used to do so.}
These properties include timezone of commit, remote URL, number of git tags in repository, name of build host, time of build, name of builder, email of builder, branch name, number of commits, local branch name, and total commits in the repository.
\newc{All of them may diverge between the builder and rebuilder.}
For example, value of \texttt{\path{git.tags}} vary for Maven release \texttt{\path{org.apache.drill:drill-opentsdb-storage:1.21.0}} due to additional number of tags in the repository during rebuild.
\textbf{Mitigation:} The rebuilder should strip out the git properties from the Jar file before comparison or avoid comparison of the Git properties.
Another solution is to fix the values for each git property.
For example, total commits should be calculated up to the commit that is used to build the artifact.
%We also reported the issue regarding this solution here~\footnote{\url{https://github.com/git-commit-id/git-commit-id-maven-plugin/issues/825}}.
See issue on our repository~\footnote{\url{https://github.com/chains-project/reproducible-central/issues/19}} to see more examples of differences in the git properties.

\subsubsection{\textbf{Unreproducible Timestamps}}

Timestamps are the most common cause of unreproducibility in our dataset as we observe them embedded in multiple ways spread across artifacts in Maven release.
They are a widely known cause of unreproducibility~\cite{bajaj_unreproducible_2023}~\cite{goswami_investigating_2020b}~\cite{benedetti_empirical_2025}.

In Maven, one solution exists to prevent timestamps from breaking reproducibility.
The Maven POM file can be configured to set the timestamp to a fixed value using property \texttt{\path{project.build.outputTimestamp}}~\footnote{\url{https://maven.apache.org/guides/mini/guide-reproducible-builds.html}}.
The value of this property is then used by Maven plugins like \texttt{\path{maven-jar-plugin}} to set the timestamp in the JAR file.

However, we observe that there are 10 other ways timestamps can be embedded in the JAR file.
In our dataset, we observe that:
1) they can be embedded in the properties file~\cref{appendix:timestamp-properties}, 
2) generated documentation~\cref{appendix:timestamp-doc}, 
3) shell scripts~\cref{appendix:timestamp-shell}, 
4) executable binaries~\cref{appendix:timestamp-binaries},
5) software bill of materials~\cref{appendix:timestamp-sbom}, 
6) JVM bytecode~\cref{appendix:timestamp-jvm}, 
7) file metadata~\cref{appendix:timestamp-files},
8) MANIFEST.MF~\cref{appendix:timestamp-manifest}, and
9) NOTICE~\cref{appendix:timestamp-notice}.
10) servlets created by Jasper JSP compiler~\cref{appendix:timestamp-servlets}.
\textbf{Mitigation:} The solution proposed by Maven is partial, more engineering is needed to support \texttt{\path{project.build.outputTimestamp}} in all Maven plugins that create the above mentioned files.

\subsubsection{\textbf{Miscellaneous Unreproducibility Reasons}}

\newc{We report reasons of reproducibility that are specific to a particular released module.
They do not fit into the categories above, so we report them here with suggested mitigations.}

\newc{\textbf{Version Ranges} Only one release \texttt{\path{org.apache.aries:org.apache.aries.jax.rs.whiteboard:2.0.2}} is observed to use version range for \texttt{flattened-maven-plugin} which affects the generation of the build manifest, pom.xml.
It is better to pin exact versions as it contributes to reproducible builds and is also a good security practice~\cite{javanjafari_dependency_2023}.}

\newc{\textbf{UUIDs:} Non-deterministic UUID generation in configuration files prevents reproducibility.}

\newc{\textbf{JavaScript Bundling:} Bundled JS files can have different variable names and line endings across builds.}
\newc{This has also been reported in the literature~\cite{goswami_investigating_2020b}.}

\newc{\textbf{Microbenchmarks:} Performance testing results should not be embedded in release artifacts.}

\newc{\textbf{Index Files:} Cache files from tools like Jandex, EFX Toolkit, and Spotless Formatter are not needed for end users.}

\subsubsection{\textbf{Mysteries}}

\newc{We observe changes in \texttt{\path{NOTICE}}, build manifests, and other configuration files that are not due to any of the above reasons and thus, we do not know what caused them and cannot propose a mitigation strategy for them.
Hence, we classify them under this section.}
\newc{We discuss the changes one by one in \cref{appendix:mysteries}.}

\newc{\subsection{Key Novelty of the Taxonomy}}

\newc{Table~\ref{tab:taxonomy} emphasizes the fine-grained causes for build unreproducibility that have never been studied in the literature before. 
Each~\cmark~indicates a novel finding in our work while empty cells represent causes that have already been reported in the literature, in particular  by Xiong et al.~\cite{xiong_build_2022}.
}

\newc{Most of the categories that we have identified are not present in the related work ~\cite{xiong_build_2022}.
The high-level reasons for unreproducibility related to SBOM and Filesystem are not mentioned at all in the related work.
Other categories are present in the work of Xiong et al., but we identify finer subcategories.
For example, we report nine more subcategories for build manifest.}

\newc{Finally, a fundamental difference between our work and Xiong et al.~\cite{xiong_build_2022} is that we perform a comparison between reference and rebuild versions of the same artifact while they build twice and compare.
We have described these fundamentally different approaches in \cref{sec:verifying-reproducibility}}.

\begin{mdframed}
  \textbf{Answer to RQ1: ``\rqtaxonomy''} \\
  We identify 6 main reasons of unreproducibility in Java and their mitigation: 1) build manifests, 2) software bill of materials, 3) filesystem, 4) JVM bytecode, 5) versioning properties, and 6) timestamps.
  \newc{We also report novelty of our findings in \autoref{tab:taxonomy} with respect to the related work.}
  %\newc{The mitigation guideline proposed provides clear guidelines to mitigate unreproducible builds in Java.} 
\end{mdframed}

\section{RQ2: Unreproducibility Mitigation with Artifact Canonicalization}
\label{sec:artifact-canonicalization}

\subsection{Objective}

The goal of this research question is to investigate if artifact canonicalization can make the build reproducible.

\subsection{Methodology}

\textbf{Artifact canonicalization} means that the entire artifact is transformed by removing non-deterministic and spurious changes, especially in metadata.
This helps downstream build tools to interpret the artifact in a consistent way.
Multiple reasons can prevent the reproducibility of an artifact(\autoref{sec:taxonomy}), such as the presence of non-deterministic information in files like MANIFEST.MF, \texttt{\path{pom.properties}}, and \texttt{\path{git.properties}} files.
Moreover, the artifact may contain non-deterministic information in the form of timestamps or file order.
Artifact canonicalization is a process of transforming the artifact into a representation that is independent of these spurious changes.
For example, the \texttt{\path{pom.properties}} file is a file generated by Maven that records the group ID, artifact ID, version, and timestamp of the build in the JAR file.
Artifact canonicalization can also remove the timestamp and set a fixed order to the declaration of properties.

\ossrebuild\footnote{\url{https://github.com/google/oss-rebuild/commit/4ef4c013fe6903cda40a9ee4244e3b65b5834325}} is a tool by Google to canonicalize software artifacts in order to improve reproducibility.
In \ossrebuild, ``canonicalizing'' is called ``stabilizing''.
It is capable of canonicalizing tarballs, ZIP files, and GZIP files. 
For example, it canonicalizes file order, modification timestamp, compression algorithm, encoding, and other miscellaneous attributes from a ZIP archive.

\chainsrebuild\footnote{\url{https://github.com/chains-project/chains-rebuild/commit/6dd67d5c7ac4db112f3419b5132d8f80a22cbe65}} is our own fork of \ossrebuild, with crucial improvements dedicated to Maven reproducibility. 
\newc{We add support for canonicalizing build manifests and embedded versioning properties in JAR files as suggested in our taxonomy (\autoref{tab:taxonomy}). }
% We strip all attributes in MANIFEST.MF which are non-deterministic and can cause unreproducibility.
\newc{For example, the \texttt{\path{Built-By}} attribute in the \texttt{\path{MANIFEST.MF}} file is sometimes updated to the username of the user who built the JAR file, a spurious change which is removed by \chainsrebuild by directly editing the byte buffer of the JAR file.}
For versioning properties, we remove the \texttt{\path{git.properties}} files embedded in the JAR file.
Finally, we remove the \texttt{\path{pom.properties}} file. 
\newc{These additional canonicalization features have been proposed to the \ossrebuild developers through 6 Pull Requests (\href{https://github.com/google/oss-rebuild/pull/339}{\#339}, \href{https://github.com/google/oss-rebuild/pull/360}{\#360}, \href{https://github.com/google/oss-rebuild/pull/363}{\#363}, \href{https://github.com/google/oss-rebuild/pull/339}{\#377}, \href{https://github.com/google/oss-rebuild/pull/414}{\#414}, and \href{https://github.com/google/oss-rebuild/pull/413}{\#413}).}

\newc{We run \ossrebuild and \chainsrebuild on the \unreproducibleartifacts reference and rebuild artifact pairs in order to analyze how effectively they make the build reproducible by canonicalization}.

\subsection{Results}

\begin{table}[h]
  \centering
  \small
  \begin{tabular}{|l|c|c|}
    \hline
    \multirow{2}{*}{\textbf{Tool}} & \textbf{Successful} & \textbf{Failed} \\
                 & \textbf{Canonicalization} & \textbf{Canonicalization} \\
    \hline
    \ossrebuild (\href{https://github.com/google/oss-rebuild/commit/4ef4c013fe6903cda40a9ee4244e3b65b5834325}{4ef4c01}) & \newc{\ossrebuildsuccess (9.41\%)} & \newc{\ossrebuildfailure (90.59\%)} \\
    \chainsrebuild (\href{https://github.com/chains-project/chains-rebuild/commit/6dd67d5c7ac4db112f3419b5132d8f80a22cbe65}{6dd67d5}) & \newc{\chainsrebuildsuccess (26.60\%)} & \newc{\chainsrebuildfailure (73.40\%)} \\
    % \jNorm + \chainsrebuild & 3303 (26.89\%) & 8980 (73.11\%) \\
    \hline
  \end{tabular}
  \caption{Effectiveness of \ossrebuild and \chainsrebuild on mitigating \unreproducibleartifacts unreproducible artifacts via canonicalization.}
  \label{tab:ossrebuild-results}
\end{table}

\autoref{tab:ossrebuild-results} shows the results of \ossrebuild and \chainsrebuild on \unreproducibleartifacts unreproducible artifacts.
The first column is the name of the tool and its commit hash we use to canonicalize the artifacts.
The second column shows the number of artifact pairs that are successfully canonicalized by \ossrebuild and \chainsrebuild.
The third column shows the number of artifact pairs that are not properly canonicalized.

% finding 1
\ossrebuild, with generic archive stabilizers, is little effective, with only 9.48\% of the artifacts becoming reproducible.
% finding 2
\chainsrebuild is significantly more effective, with 2.5x more success, thanks to dedicated features for canonicalizing JAR files.
Recall that, we add support for canonicalizing MANIFEST, embedded versioning properties, and pom.properties files in JAR files.
For example, artifact \texttt{\path{slf4j-ext-2.0.6.jar}} of Maven release \texttt{\path{org.slf4j:slf4j-ext:2.0.6}} is successfully canonicalized by \chainsrebuild.
Originally, there are differences in zip metadata and MANIFEST.MF.
\ossrebuild only canonicalizes the zip metadata. \chainsrebuild takes  a step further and canonicalizes the MANIFEST.MF file as well.
It fixes the ordering of values in MANIFEST.MF under the \texttt{\path{Export-Package}} attribute.
As a minified example, the fix changes \texttt{\path{Export-Package: org.slf4j.ext;version="2.0.6",org.slf4j.agent;version="2.0.6"}} to \texttt{\path{Export-Package: org.slf4j.agent}}.
Here arranging the order of the values in \texttt{\path{Export-Package}} attribute leads to a canonicalized output and a reproducible build.

% concept
Overall, this experiment is evidence that artifact canonicalization is a valid solution to mitigate some unreproducible builds.
However, it is only a partial workaround and does not replace the mitigation fixing build and rebuild process (see \autoref{sec:mitigation}).
As shown in \autoref{tab:taxonomy}, there are many other causes of unreproducibility that can be fixed by canonicalization.
For example, non-deterministic information in SBOMs and environment influences in files can be fixed by canonicalization.
This includes SBOM, generated JVM bytecode, and ordering of sections in JVM bytecode.
We leave this as future work as it is purely an engineering task.

\begin{mdframed}
  \textbf{Answer to RQ2: ``\rqossrebuild''} \\
  \newc{Our publicly available prototype \chainsrebuild can canonicalize (\chainsrebuildsuccess) 26.60\% of the artifacts from our dataset.}
  \newc{Our study is the first large-scale experiment on artifact canonicalization, demonstrating that it is a promising mitigation of build unreproducibility.}.
\end{mdframed}

\section{RQ3: Unreproducibility Mitigation with Bytecode Canonicalization}
\label{sec:bytecode-canonicalization}

\subsection{Objective}

In \autoref{sec:taxonomy}, we observe that changes in JVM bytecode is one of the causes of unreproducible builds.
In this research question, we investigate to what extent these changes can be mitigated by using bytecode canonicalization.

\subsection{Methodology}

\textbf{Bytecode canonicalization} is a process of transforming the bytecode of a program into a representation that is independent of specific implementation details inserted by the compiler.
This is either done by converting the bytecode to an intermediate form  or by rewriting the bytecode.
For example, subtraction of two integers is implemented as addition of a positive and negative integer in Java 6 and higher versions.
Meanwhile, in Java 5 and lower versions, it is implemented as subtraction of two integers.
Hence, if the program is compiled with Java 5 and 6, the two bytecode  representations differ even though the program is semantically the same --- performing subtraction of two integers.
Canonicalization removes these differences by converting the bytecode to an intermediate form which is free from the implementation details that vary across different versions of the compiler.

We consider the state-of-the-art Java bytecode canonicalization tool by Schott et al.~\cite{schott_java_2024}, called \jNorm. 
\jNorm's primary goal is to remove the parts from Java bytecode that may differ due to changes in JDK version and output the canonicalized version of the bytecode.
\jNorm only removes the differences that do not change the semantic behavior of the artifact.
Note that they refer to ``canonicalization'' as ``normalization'' in their tool.

We run \jNorm version \texttt{\path{1.0.0}} on the \unreproducibleartifactswithjvmbytecode reference rebuild artifact pairs that have differences in the bytecode.
For maximum canonicalization, we run the tool with all the command line options as listed in the README~\footnote{\jNorm tool --- \url{https://github.com/stschott/jnorm-tool/tree/cec4645c5c9b52f73c347349bf14945b0eb55c87}} of the repository and tool's help output.
As a result of canonicalizing the artifacts, jNorm produces files in the Jimple~\cite{rajavallee-rai_Jimple_1998} format. We save the canonicalized Jimple files for both the reference and the rebuild artifacts.
There are cases where \jNorm crashes and canonicalization fails, which we consider as a failure.
We do not analyze them further as they are bugs in tool, but we upload the logs for future debugging.
We run the Linux diff tool over the canonicalized pairs of Jimple files.
If there is no difference, it means that canonicalization is a success, for reproducibility.
In other words, successful canonicalization by \jNorm produces two identical artifacts.
We report the number of cases where the diff between the pair is empty.
Finally, we analyze the cases where \jNorm is unable to fully canonicalize the bytecode in order to understand its fundamental limitations and identify required future work.

\subsection{Results}
% grade daleq
% Successful normalization: 5
% Failed normalization: 33
% Error in normalization: 0

\newc{Out of \unreproducibleartifactswithjvmbytecode artifacts, \jNorm's execution leads to the following distribution: it successfully canonicalizes \jnormsuccess (29.7\%) artifacts which means it produces two identical artifacts; it leaves \jnormfailure (53.2\%) artifacts with a diff after canonicalization; and there are \jnormerror (17.1\%) crashes, which are ignored.}

Our results are in contrast with the results of Schott et al.~\cite{schott_java_2024} where the authors report that \jNorm is able to canonicalize up to 99\% of pairs of Java classfiles when JDK version differs.
However, our success ratio is lower (29.7\%).
This can be attributed to the different methodology used to generate the dataset.
\jNorm is evaluated on Java classfiles built on the same machine but varying JDK versions.
In our case, the reference and rebuild artifacts are built on different machines and based on our results in \autoref{sec:taxonomy}, we know that the build process is different.

Next, we study the reasons why \jNorm fails to produce identical artifacts when run on \jnormfailure cases.
We categorize them based on the features in the bytecode that are not properly canonicalized by \jNorm.
Note that some unreproducible artifacts may fall into multiple categories as it is possible \jNorm is unable to canonicalize multiple problems in the same artifact.

\paragraph{Structural Change Limitation}

Recall from \autoref{sec:jvm-bytecode} that the order of methods, fields, static initializers, entries in the constant pool, and array type values in annotation sometimes differ between the reference and rebuilt artifacts.
We find that \jNorm is unable to canonicalize most of those problems: the order of fields, methods, static initializers, array type values in annotation, removal of implicit modifier from fields.
The only mitigated problem by \jNorm is the constant pool order problem, because the Jimple format abstracts away the constant pool~\cite{vallee-rai_soot_1999}.
We now discuss concrete examples of failed canonicalization:
\begin{itemize}
\item  \texttt{\path{org.apache.shiro:shiro-aspectj:1.11.0}} is not canonicalized due to the changed order of static initializers in the bytecode.
\item  \texttt{\path{antisamy-1.7.3.jar}} has a difference in implicit visibility modifier between the reference and rebuild artifact, not handled by \jNorm.
\item  \texttt{\path{org.apache.helix:zookeeper-api:1.0.3}} has one unreproducible artifact which is due to the naming of lambda functions in the bytecode.
Since the names of lambda functions are suffixed by integers determined non-deterministically by the compiler, \jNorm is unable to canonicalize them.
\end{itemize}

% Lambda functions are assigned different numbers in reference and rebuild artifacts and \jNorm is not able to canonicalize them.
% We observe this in \texttt{\path{zookeeper-api-1.0.3.jar}} artifact.
\paragraph{Control Flow Limitation}
There are two cases of this category where \jNorm is unable to canonicalize changes in control flow.
\begin{itemize}

\item The first case is change in control flow of if statements as illustrated in \autoref{lst:javap-diff} and \autoref{lst:jnorm-diff}. \texttt{\path{eu.maveniverse.maven.mima:2.4.2:standalone-static-uber}} have changes in the control flow of if conditions. The reference checks for the negative condition of the if statement while the rebuild checks for the positive condition first, because the two builds use different compilers. This is shown in the diff generated by \diffoscope in \autoref{lst:javap-diff}. \jNorm does not canonicalize control flow there is still a diff in Jimple format, as seen in \autoref{lst:jnorm-diff}.

\begin{figure}[h]
  \begin{minipage}[t]{.48\columnwidth}
    \begin{lstlisting}[caption={javap diff for control flow difference.}, style=diff, label={lst:javap-diff}]
    %\RHilight%-   99: ifne       106
    %\RHilight%-  102: iconst_1
    %\RHilight%-  103: goto       107
    %\RHilight%-  106: iconst_0
    %\RHilight%-  107: ireturn
    %\GHilight%+   99: ifeq       104
    %\GHilight%+  102: iconst_0
    %\GHilight%+  103: ireturn
    %\GHilight%+  104: iconst_1
    %\GHilight%+  105: ireturn
    \end{lstlisting}
  \end{minipage}\hfill
  \begin{minipage}[t]{.48\columnwidth}
    \begin{lstlisting}[caption={\jNorm diff for control flow difference.}, style=diff, label={lst:jnorm-diff}]
    %\RHilight%-if v != 0 goto label;
    %\RHilight%-v = 1;
    %\RHilight%-goto label;
    %\RHilight%-label:
    %\RHilight%-v = 0;
    %\GHilight%+if v == 0 goto label;
    %\GHilight%+return 0;
    label:
    %\RHilight%-return v;
    %\GHilight%+return 1;
    \end{lstlisting}
  \end{minipage}
\end{figure}

\item Second, Maven release \texttt{\path{io.fabric8:kubernetes-model-apiextensions:6.4.0}} contains generated code where the order of if-else blocks is different across different builds. This is not an issue with \jNorm, this is an issue with the unreproducibility of the code generator employed during build.

\end{itemize}

\paragraph{Embedded Data Limitation}
In \autoref{sec:taxonomy}, we discussed that absolute file paths, timestamps, Java version, and project version are embedded in the bytecode.
\jNorm is able to canonicalize the Java version and project version as it deletes the \texttt{\path{module-info.class}} file that stores this information.
However, it is unable to canonicalize absolute file paths and timestamps\footnote{\jNorm is able to canonicalize timestamp if they are declared in a static final field.}.

\paragraph{Optimization Limitation}
We find 5 cases of optimizations or de-optimizations that are not canonicalized by \jNorm.
We present the different features in Java that are not canonicalized by \jNorm.
For a more detailed discussion, refer to the ~\autoref{appendix:jnorm-optimizations-failure}.
\begin{itemize}
  \item JDK 17 creates a lookup table to store the values of the enum while JDK 22 simply stores it based on the order of the enum values.
  \item JDK 17 uses the \texttt{\path{invokevirtual}} bytecode instruction when handling invocations to \texttt{\path{toString}}, \texttt{\path{equals}}, and \texttt{\path{getClass}}, while JDK 21 uses \texttt{\path{invokeinterface}}. This is due to a change proposed in JDK 18~\footnote{\url{https://github.com/openjdk/jdk/pull/5165}}.
  \item String concatenation expressions where one of the operands requires a cast to String.
  \item Implementation of try and try-with-resource statements and finally clauses vary between Java 5, 8, and 11.
  \item Reference to outer class from inner class is handled differently in patch versions of Java 17.
\end{itemize}

We do not delegate fixing all of these problems to \jNorm.
Recall that we mentioned in \autoref{sec:taxonomy} that rebuilder can either fix the build process or canonicalize the output in order to mitigate the unreproducibility.
\jNorm can fix problems in bytecode structure, control flow, and embedded data.
However, unreproducible optimization and de-optimization can be mitigated by fixing the Java version in the build process.

We note that multiple canonicalization tools can be used in conjunction.
For instance, \chainsrebuild and \jNorm can be used together to canonicalize the JVM bytecode and the JAR files of a Maven artifact.
Since \chainsrebuild can canonicalize \chainsrebuildsuccess artifacts and \jNorm can canonicalize \jnormsuccess artifacts, we can combine the 28.74\% (3,680 / \unreproducibleartifacts).
% \todo{compute combined releases} \newc{Out of 3,680 artifacts, 2,578 + 361 artifacts are canonicalized such that 2,263 + 230 / \unreproduciblereleases (28.43\%) releases become fully reproducible.}
\newc{Out of 3,680 artifacts, 2,939 artifacts are canonicalized such that \canonicalizedreleases / \unreproduciblereleases (30.06\%) releases become fully reproducible.}

\begin{mdframed}
  \textbf{Answer to RQ3: ``\rqjnorm''} \\
  \newc{\jNorm is able to canonicalize \jnormsuccess (29.38\%) out of \unreproducibleartifactswithjvmbytecode bytecode artifacts.
  However, there are \jnormfailure (54.27\%) artifacts that are still unreproducible after canonicalization.}
  We present in detail four limitation of bytecode canonicalization related to structural change, control flow, embedded data, and optimization.
  \newc{Finally, our experiments demonstrate that combining artifact and bytecode canonicalization successfully mitigates unreproducibility in 28.74\% (3,680 / \unreproducibleartifacts) artifacts.}
\end{mdframed}

\section{Best Practices for Builders and Rebuilders}

\subsection{Builders}
Builders can improve reproducibility by ensuring deterministic build inputs, such as pinning dependency versions (), clearing caches before deployment, and using standard build commands rather than ad hoc scripts.
They should avoid embedding non-deterministic data like branch names, timestamps, usernames, or UUIDs in manifests and properties, and instead use plugins that enforce deterministic ordering.
To support verification, builders should also document and publish the exact JDK version and vendor used for releases, and generate SBOMs in a consistent and reproducible manner.

\subsection{Rebuilders}
Rebuilders should strive to mirror builder environments as closely as possible by using the same JDK vendor and version, along with consistent settings for locale, time zone, and operating system.
When unreproducibility persists, they can apply canonicalization tools such as \ossrebuild and \jNorm to remove spurious differences in bytecode and artifacts.
These tools are carefully crafted to be able to canonicalize unreproducibility issues in the bytecode and artifacts.
The remanining diff after first canonicalization can help as a feedback to the builder to fix the unreproducibility issue or to write more canonicalization rules.

\section{Related Work}
\label{sec:related-work}

\subsection{Reproducibility in Applications}
Reproducibility is a key feature in state of the art systems.
One of the first software to become reproducible was Bitcoin~\footnote{\url{https://github.com/bitcoin/bitcoin}} ensuring that none of the nodes in the Bitcoin peer-to-peer network contains backdoors.
Tor Browser\footnote{\url{https://gitlab.torproject.org/tpo/core/tor}}, a web browser that anonymizes the user's web traffic, is also reproducible since 2013~\cite{levsen_tale_2025}.
Tails, an operating system that protects against surveillance and censorship, also announced in 2017 that their ISO images are reproducible~\footnote{\url{https://tails.net/news/reproducible_Tails/}}.
Both the Tor Browser and Tails operating system emphasize that Reproducible Builds are necessary for software geared towards privacy and security.

The Reproducible Builds~\footnote{\url{https://reproducible-builds.org/}} project works on the reproducibility of coreboot~\footnote{\url{https://tests.reproducible-builds.org/coreboot/coreboot.html}} --- an open-source firmware for x86 and ARM systems, Debian~\footnote{\url{https://tests.reproducible-builds.org/debian/reproducible.html}} --- a Linux distribution, FreeBSD~\footnote{\url{https://tests.reproducible-builds.org/freebsd/freebsd.html}} and NetBSD --- a UNIX based operating systems, and OpenWrt~\footnote{\url{https://tests.reproducible-builds.org/openwrt/openwrt.html}} --- operating system for embedded devices.

In the Java ecosystem, Eclipse Temurin, a vendor for Java Developement Kit, provides reproducible builds for their JDK~\cite{leonard_eclipse_2024}.
The build of Apache Maven is also reproducible~\footnote{\url{https://github.com/jvm-repo-rebuild/reproducible-central/blob/master/content/org/apache/maven/maven/README.md}}.
Our work aims to enable ever more Java software applications to become reproducible by providing comprehensive and effective mitigation strategies.

\subsection{Verifying Reproducibility}
\label{sec:verifying-reproducibility}

In the literature, we have found three ways to check reproducibility.
First approach is to build once and compare the generated artifacts with the artifacts on package registry.
Second is to build twice with variations in environment and compare the generated artifacts.
Finally, third approach is to verify that the projects stand the test of time by ensuring that they build successfully over time.

\paragraph{Comparison between reference and rebuild artifacts}
\label{sec:rw-comparison}
One way to verify build reproducibility is to  build once and compare the generated artifacts with the artifacts on package registry. 
This is what we do in this paper.

Outside the Java ecosysem, 
Malka et al.~\cite{malka_does_2025} investigate reproducibility in functional package manager Nix.
The paper proves that functional package managers help reproduciblity by showing 91\% of the packages in Nix are reproducible.
Similarly, Bajaj et al.~\cite{bajaj_unreproducible_2023} investigate reproduciblity in Debian and Arch Linux packages.
Apart from uncovering the causes of unreproducibility, they also do a survival analysis of packages to see how long it takes to fix the unreproducibility and how long the package stays reproducible.
They show that Arch Linux packages become reproducible sooner than Debian packages.
However, a Debian package is likely to stay reproducible for longer before it becomes unreproducible again.
Moritz~\cite{moritz_decentralized_2023} proposes a decentralized protocol on Hyperledger Fabric to verify reproducibility.
The protocol involves comparing the locally generated buildinfo file with the one on Debian registry.
Linderud~\cite{linderud_reproducible_2019} also proposes a protocol to verify buildinfo files, but a layer of transparency log is added in order to ensure reproducibility status for specific builds are immutable.
Drexel et al.~\cite{drexel_reproducible_2024} contributes with an independent rebuilder that verifies reproducibility of Arch Linux packages based on this approach.
Vu et al.~\cite{vu_lastpymile_2021} focus on finding the most common files and APIs that differ between source repository and Python wheel published on PyPI.
Goswami et al.~\cite{goswami_investigating_2020b} investigate reproducibility in NPM packages.
They compare the minified artifacts generated by JavaScript bundlers to report the reasons of unreproducibility.
Although there are overlap in taxonomies reported by these works, our analysis focuses on Java ecosystem and proposes novel reasons for unreproducibility.
For example, variations in MANIFEST, JVM bytecode, and versioning properties are first seen in our work.
Even for reasons such as timestamps, we are the first to report 9 ways that it can be embedded in a Maven release.

Some works analyze single projects rather than the entire ecosystems.
Shi et al.~\cite{shi_experience_2022a} investigate reproducibility for four commercial systems running in Huawei and CentOS --- a production grade Linux operating system.
Pöll et al.~\cite{poll_automating_2022} discover reproducibility issues while rebuilding Android 5 to 12 images.
Carné de Carnavalet et al.~\cite{decarnedecarnavalet_challenges_2014} investigate reproducibility for 16 versions of the TrueCrypt encryption tool.
Finally, Andersson~\cite{andersson_geth_2024} investigates reproducibility for Go Ethereum binary.
None of the subjects in these related work are Java based.
The work of Pöll et al.~\cite{poll_automating_2022} is Java based, but Android images do not compile to JVM bytecode like traditional Java applications.
Hence, ours paper is the first one to investigate reproducibility of normal Java applications at scale.

\paragraph{Comparison between two subsequent builds}

Benedetti et al.~\cite{benedetti_empirical_2025} verify the reproducibility of 12,000 Ruby, PyPI, and Maven packages using this approach.
They keep all inputs same except the environment which is changed for each build using \texttt{\path{reprotest}}~\footnote{\url{https://salsa.debian.org/reproducible-builds/reprotest}}.
\texttt{\path{reprotest}} can change locale, timezone, and number of CPUs for example.
Ren et al.~\cite{ren_automated_2018b,ren_automated_2022} and Leija et al.~\cite{navarroleija_reproducible_2020} propose a technique to localize and fix the root cause of unreproducibility in Java artifacts.
They build the source files and dependencies for Debian packages twice and compare the artifacts using \diffoscope~\footnote{\url{https://diffoscope.org/}}.
Lagnöhed~\cite{lagnohed_integration_2024} propose a tool \texttt{\path{rmake}} built on top of \texttt{\path{make}} with support for building projects twice and comparing the artifacts.

\newc{In Java, the most closely related work here is the work by Xiong et al.~\cite{xiong_build_2022}. They compare 59 Maven projects on Maven Central and internal registry of Huawei with the artifacts generated by the build process.
They discover build reproducibility problems due to environment, JDK, multithreading, and other tools.
% jq '.[].name' java/unreproducible_maven_projects_to_releases.json | cut -d':' -f1-2 | sort | uniq | wc -l = 263 Maven projects
Our work differs in the scale of the dataset (ours is 4$\times$ larger) and the taxonomy of the root causes of unreproducibility.
Our taxonomy has more categories than the one proposed by Xiong et al. including new essential categories such as variations in filesystem and software bill of materials.}

All of these works differ from our work in the approach they take to verify reproducibility.
We claim that the first approach uncovers more reasons of unreproducibility as it compares artifacts generated by build process which have maximum differences in environment.
Carné de Carnavalet et al.~\cite{decarnedecarnavalet_challenges_2014} also emphasizes that there should be maximum difference between the environments to detect unreproducibility.

\paragraph{Build Outcome Analysis}

Build success is a prerequisite to study reproducibility. Maes-Bermejo et al.~\cite{maes-bermejo_revisiting_2022}, Hassan et al.~\cite{hassan_automatic_2017}, \cite{sulir_quantitative_2016}  and Yasi et al.~\cite{yasi_automatic_2022} show that only 40-60\% of the Maven projects in their respective datasets build successfully.
This indicates that around half of the projects don't even generate artifacts as their build process fails.
All of these works are different from ours as they focus on the success of the build process only  and do not look at the generated artifacts.
Our work focuses on artifacts to identify the reasons of unreproducibility.
Technically, Maes-Bermejo et al.~\cite{maes-bermejo_revisiting_2022} use \texttt{\path{mvn clean compile -X}} to build all the Maven projects.
In our work, our build commands in the dataset are tailored to each project so that the build process succeeds.

% 52-yasi fail to build (default)
% 58 - maes
% 38 - sulir (default)
% 38 - hassan (default)

\subsection{Fixing Unreproducibility}
\label{sec:fixing-unreproducibility}

We now review the related work on fixing causes of unreproducibility.

Ren et al.~\cite{ren_automated_2022, ren_root_2019, ren_automated_2018b} propose techniques to localize and fix unreproducibility in Debian packages.
They localize the file in the Debian package that causes unreproducibility and then suggest a patch based on a history of patches for fixing unreproducible builds.
This approach is specific for Debian packages and does not apply to Java artifacts.
Moreover, their approach to detect unreproducibility is different from ours as mentioned in~\cref{sec:rw-comparison}.

Mukherjee et al.~\cite{mukherjee_fixing_2021b} propose a technique to fix unreproducible Python builds caused by dependency errors.
They report that 71\% of Python builds in their dataset fail; they fix these builds by analyzing the logs.
This work is only comparess success of builds, not artifacts, and it is focussed on the Python ecosystem.

Xiong et al.~\cite{xiong_build_2022} classifies fixing unreproducibility into three categories --- 1) remediation by patching source code, third-party dependency, or build script 2) controlling which means interception of non-deterministic build instructions and returning pre-defined values for them 3) interpretation which means documenting the reason for unreproducibility.
The authors perform these fixes automatically using a tool called \texttt{\path{JavaBEPFix}} which is not available to the public.
Hence, we cannot evaluate the effectiveness of their tool on our dataset.

Keshani et al.~\cite{keshani_aroma_2024} automates the process of reproducing Maven projects by generated buildspec files.
They show that their approach simplifies generated buildspec files and also fixes issues in the currently existing buildspec files in Reproducible Central dataset.
Our work also finds that a mitigation strategy to fix unreproducibility issue is by fixing the build script. In addition, we perform the first analysis of fixing by canonicalization.

Randrianaina et al.~\cite{randrianaina_options_2024} identifies the configuration options in that cause unreproducibility.
The authors select highly configurable systems such as \texttt{\path{Linux}}, \texttt{\path{BusyBox}}, and \texttt{\path{ToyBox}} to study the unreproducibility of their builds.
They find options for module signing, debug information, and profiling that cause unreproducibility.
In our work, configuration options is one of the causes of unreproducibility.
Especially, when the build process involves setting Maven profiles~\autoref{sec:jvm-bytecode}. 

Dietrich et al.~\cite{dietrich_levels_2024} show that only a 25\% (out of 14,156) of the binaries in Maven ecosystem are actually bit by bit identical.
These binaries come from four different registries --- Maven, Google Assured Open Source Software, RedHat, and Oracle build-from-source.
They propose four levels of binary equivalence to compare binaries which becomes increasingly lenient as we go from level 1 to level 4.
First is to compare binaries bit by bit and this is the strictest level of binary equivalence check.
Second is to compare binaries and ignore differences that have no semantic effect (for example, \texttt{\path{@Deprecated}}).
\ossrebuild fits into the second level of checking binary equivalence.
Third is to convert binaries into a common format and then compare them.
\jNorm fits into the third level of checking binary equivalence
Finally, fourth is to compare binaries using a similarity metric.
In the paper, they use \texttt{\path{tlsh}}~\cite{oliver_tlsh_2013} to compare hash distance between two binaries; lesser the hash distance, more similar the binaries are.
This work is similar to ours as they also evaluate effectiveness of \jNorm over thousands of pairs of JVM bytecode~\cite{dietrich_bineqa_2024} to fix unreproducibility issues.
However, we consider more causes of unreproducibility other than ones in JVM bytecode.

Finally, \texttt{\path{strip-nondeterminism}}~\footnote{\url{https://salsa.debian.org/reproducible-builds/strip-nondeterminism}} is a tool the Debian project that strips out timestamps and file system order from archives.
However, it only takes care of these two causes of unreproducibility and does not cover the other causes of unreproducibility that we have discussed in this paper.

\section{Conclusion}

In this paper, we have presented the first large-scale study of build unreproducibility in Java. We have identified the six main causes of unreproducibility with underlying root causes: 1) build manifests, 2) software bill of materials, 3) filesystem, 4) JVM bytecode, 5) versioning properties, and 6) timestamps.
Leveraging the same dataset, we have evaluated the effectiveness of canonicalization as a mitigation strategy to fix unreproducibility. Our results have shown that 26.89\% of unreproducible artifacts can be canonicalized and become reproducible after canonicalization.

As future work, we want to formalize the notion of acceptable canonicalization by verifying what transformations can be applied to build artifacts without obscuring meaningful differences.
This is important to ensure that the technique eliminates non-deterministic noise while preserving semantic integrity. Build reproducibility verification via canonicalization must be sound and precise.

\section{Acknowledgements}

We acknowledge the use of GitHub Copilot for assistance in polishing the writing of this paper.
This work is supported by the CHAINS project funded by Swedish Foundation for
Strategic Research (SSF), as well as by the Wallenberg Autonomous Systems and Software Program (WASP) funded by the Knut and Alice Wallenberg Foundation.

\bibliographystyle{ACM-Reference-Format}
\bibliography{main}

\pagebreak

\appendix

\section{Unreproducible MANIFEST.MF}

The MANIFEST.MF file is a mandatory metadata file that contains information about the JAR file.

\subsection{\texttt{Built-By}}
\label{appendix:mf-built-by}

This attribute is used to indicate the name of the user who built the JAR file.
This can be different for environments as the user name is different.

\textbf{Example} Maven release \texttt{\path{org.apache.camel:came-ahc-ws:3.13.0}} has an unreproducible artifact \texttt{\path{camel-ahc-ws-3.13.0.jar}}.
This JAR file has a different \texttt{\path{Built-By}} attribute in the MANIFEST.MF file.
In the reference version of the JAR file, the \texttt{\path{Built-By}} attribute is set to \texttt{\path{root}} while in the rebuild version of the JAR file, the \texttt{\path{Built-By}} attribute is set to \texttt{\path{aman}}.

\textbf{Solution} Remove the \texttt{\path{Built-By}} attribute from the MANIFEST.MF file.

\subsection{\texttt{Implementation-Build-Java-Vendor}}
\label{appendix:mf-implementation-build-java-vendor}

This attribute records the vendor of the JDK used to build the JAR file.
There are 17 different open source distributions of JDK~\footnote{\url{https://sdkman.io/jdks/}} --- GraalVM, OpenJDK, Temurin to name a few.
Although all of them follow the same Java specification by Oracle, the implementation details could differ.

\textbf{Example} MANIFEST.MF of \texttt{\path{ldapchai-0.8.0.jar}} of Maven release \texttt{\path{com.github.ldapchai:ldapchai:0.8.0}} shows difference in vendor when rebuilt.
It changes from \texttt{\path{AdoptOpenJDK}} to \texttt{\path{Oracle Corporation}}.
\texttt{\path{AdoptOpenJDK}}, now known as \texttt{\path{Adoptium}}, indicates that the JAR file is built using \texttt{\path{Temurin}}.
We use \texttt{\path{OpenJDK}} for the rebuild version, so the vendor is set to \texttt{\path{Oracle Corporation}} as \texttt{\path{OpenJDK}} is maintained by Oracle.

\textbf{Solution} The \texttt{\path{buildspec}} file should document the exact vendor of the JDK.

\subsection{\texttt{Build-Jdk} or \texttt{Implementation-Build-Java-Version}}
\label{appendix:mf-build-jdk}

This attribute reports the exact version of JDK used to build the JAR file.
The JDK version given a fixed vendor is based on Semantic Versioning.
Ensuring the exact same JDK version across different environments depends upon the major, minor, patch version and if this version has a pre-release flag or not.
These four factors are in addition to the vendor of the JDK as discussed above.

\textbf{Example} Artifact \texttt{\path{camel-cloud-3.13.0.jar}} of Maven release \texttt{\path{org.apache.camel:camel-cloud:3.13.0}} is built with different versions of JDK.
The reference version of the JAR file has the \texttt{\path{Build-Jdk}} attribute set to \texttt{\path{1.8.0_292}} while the rebuild version of the JAR file has the \texttt{\path{Build-Jdk}} attribute set to \texttt{\path{1.8.0_422}}.
The difference in the JDK version is due to the update number~\cite{clark_jep_2014}.

\textbf{Solution} The \texttt{\path{buildspec}} file should document the exact version of JDK and provide an automated way to rebuild.
There are some cases where the JDK version is precisely mentioned in Reproducible Central dataset.
However, we remove them in our dataset filtering process as they require manual intervention to build.

\subsection{Signed JARs}
\label{appendix:mf-signed-jars}

Java provides a way to sign each file in a JAR to ensure its users that the file has not been tampered with~\footnote{\url{https://docs.oracle.com/javase/tutorial/deployment/jar/intro.html}}.

\textbf{Example} The JAR file \texttt{\path{xmlchai-0.1.0.jar}} of Maven release \texttt{\path{org.jrivard.xmlchai:xmlchai:0.1.0}} shows that signatures are removed in the rebuild version of the JAR file.

\textbf{Solution} Strip the signature from the JAR file as signatures can not be replicated.

\subsection{\texttt{Os-Version}}
\label{appendix:mf-os-version}

This attribute is used to indicate the operating system version used to build the JAR file.

\textbf{Example} The artifact \texttt{\path{jandex-maven-plugin-3.1.0.jar}} from Maven release \texttt{\path{io.smallrye:jandex-maven-plugin:3.1.0}} has different \texttt{\path{Os-Version}} attribute in the MANIFEST.MF file.
It seems that the kernel version is different as the value changes from \texttt{\path{5.15.0-1035-azure}} to \texttt{\path{5.15.0-118-generic}}.

\textbf{Solution} This attribute should be stripped from the MANIFEST.MF file before comparison as it is environment specific and Reproducible Builds are environment independent.

% both are non-standard attributes and can have anything
\subsection{\texttt{Created-By}}
\label{appendix:mf-created-by}

These attribute is used to indicate the name and version of the build tool or the Java version used to produce the JAR file.
This can be created by Java \texttt{\path{jar}} tool or any build tool like Maven, Gradle, or Ant.

\textbf{Example} \texttt{\path{jline-console-3.22.0-sources.jar}} of Maven release \texttt{\path{org.jline:jline-console:3.22.0}} has different \texttt{\path{Created-By}} attribute in the MANIFEST.MF file.
The reference version reports \texttt{\path{Apache Maven}} while the rebuild version reports \texttt{\path{Maven Source Plugin 3.2.1}}.

\textbf{Solution} The \texttt{\path{buildspec}} file should document the exact variant and version of the build tool used to build the JAR file as this ensures the same build tool is used across different environments.

\subsection{\texttt{Originally-Created-By}}
\label{appendix:mf-originally-created-by}

This attribute is added by \texttt{\path{maven-bundle-plugin}} when the JAR file is created again without cleaning the build directory.
The plugin updates the JAR with this attribute to indicate the original build plugin that created the JAR file.

\textbf{Example} Maven release {ch.qos.logback:logback-access:1.3.0-alpha14} has an artifact \texttt{\path{logback-access-1.3.0-alpha14.jar}} which does not have the \texttt{\path{Originally-Created-By}} when rebuilt.

\textbf{Solution} The \texttt{\path{buildspec}} file should document prerequisites to build the JAR file for the same reasons as \texttt{\path{Created-By}} attribute.

\subsection{\texttt{SCM-Revision}}
\label{appendix:mf-scm-revision}

MANIFEST file can contain information related to version control system like Git.
\texttt{\path{SCM-Revision}} attribute is used to indicate the commit hash of the source code used to build the JAR file~\footnote{\url{https://www.mojohaus.org/buildnumber-maven-plugin/usage.html}}.

\textbf{Example} The JAR file \texttt{\path{io.wcm.tooling.commons.crx-packmgr-helper-2.0.2.jar}} of Maven release \texttt{\path{io.wcm.tooling.commons:io.wcm.tooling.commons.crx-packmgr-helper:2.0.2}} has different \texttt{\path{SCM-Revision}} attribute in the MANIFEST.MF file.
The value changed from \texttt{\path{c48c286d at 2022-01-06T11:46:12+01:00}} to \texttt{\path{bc9cc174 at 2024-10-17T02:14:22Z}}.
There are two reasons of differences here.
First, the commit hash is different which indicates that commit hash corresponding to the git tag has changed.
Second, the timestamp is different and this is due to embedding build time in the same attribute.

\textbf{Solution} The commit hash should remain the same across different builds as it guarantees that the source code is the same across different builds.
\texttt{\path{buildspec}} files currently relies on Git tags to fetch source code, but they are mutable and can be moved to different commits.
Thus, the \texttt{\path{buildspec}} file should document the exact commit hash of the source code.

\subsection{\texttt{SCM-Git-Branch}}
\label{appendix:mf-scm-git-branch}

This attribute is used to indicate the branch of the source code used to build the JAR file.

% this is the only instance in our dataset.
\textbf{Example} MANIFEST.MF of \texttt{\path{ldapchai-0.8.0.jar}} of Maven release \texttt{\path{com.github.ldapchai:ldapchai:0.8.0}} shows that the value changes from \texttt{\path{master}} to \texttt{\path{338023a44e4dc62aff8985ca42c2f6743258b1c0}}.

\textbf{Solution} Branches are mutable in Git so they should not be embedded in the MANIFEST.MF file.
Commit hash should be used instead as they are fixed.

\subsection{\texttt{Bnd-LastModified}}
\label{appendix:mf-bnd-lastmodified}

This attribute is part of the OSGi specification to create bundles.

\textbf{Example} \texttt{\path{shiro-cas-1.9.0.jar}} is an artifact of \texttt{\path{org.apache.shiro:shiro-cas:1.9.0}} where the value of this attribute changes from \texttt{\path{1647431421514}} to \texttt{\path{1735860648961}}.

\textbf{Solution} Apache Maven provides a property \texttt{\path{${project.build.outputTimestamp}}} that fixes the timestamp and can be read from again in subsequent rebuilds~\footnote{\url{https://maven.apache.org/guides/mini/guide-reproducible-builds.html}}.
This property should can either be reused to set the value of \texttt{\path{Bnd-LastModified}} attribute or the attribute should be stripped from the MANIFEST.MF file.

\subsection{Order of values for attributes}
\label{appendix:mf-order}

If the value of an attribute is a comma separated list, the order of the values can differ across different builds.

\textbf{Example} \texttt{\path{slf4j-api-2.0.6.jar}} of Maven release \texttt{\path{org.slf4j:slf4j-api:2.0.6}} has an attribute \texttt{\path{Export-Package}} in the MANIFEST.MF file.
This attribute is part of OSGi specification and is used to export packages that can be imported by other bundles.
The value of this attribute is a list of packages whose order differs across the different builds.

\textbf{Solution} The attributes where this problem exists are \texttt{\path{Include-Resource}}, \texttt{\path{Private-Package}}, and \texttt{\path{ Provide-Capability}}.
The solution is to depend upon the latest version of \texttt{\path{maven-bundle-plugin}} as it makes the order deterministic by sorting the values~\footnote{https://github.com/apache/felix-dev/commit/d885d99a6a16660f655a4fd18e8a1a39beef0a15}.

% there is another example \texttt{\path{Main-Class}}
% +Originally-Created-By:
\subsection{Addition or removal of attributes}
\label{appendix:mf-add-remove}

Presence or absence of an attribute in the MANIFEST.MF file can also cause unreproducibility.

\textbf{Example} \texttt{\path{apache-any23-csvutils-2.7-sources.jar}} of Maven release \texttt{\path{org.apache.any23:apache-any23-csvutils:2.7}} removed 10 attributes from the MANIFEST.MF file.
These attributes are \texttt{\path{Specification-Title}}, \texttt{\path{Specification-Version}}, \texttt{\path{Specification-Vendor}}, \texttt{\path{Implementation-Title}}, \texttt{\path{Implementation-Version}}, \texttt{\path{Implementation-Vendor}}, \texttt{\path{Implementation-Build}}, \texttt{\path{Implementation-Build-Date}}, \texttt{\path{X-Compile-Source-JDK}}, and \texttt{\path{X-Compile-Target-JDK}}.

\textbf{Solution} We suggest to canonicalize these attributes. We cannot confirm what build tool embeds these attributes.

\section{Unreproducible pom.properties}

The \texttt{\path{pom.properties}} file is an automatically generated file that contains information about the Maven project.

\subsection{Order of properties}
\label{appendix:pom-properties-order}

The file contains properties \texttt{\path{groupId}}, \texttt{\path{artifactId}}, \texttt{\path{version}}.
However, their order can differ across different builds.

\textbf{Example} The \texttt{\path{pom.properties}} file of Maven release \texttt{\path{io.dropwizard.metrics:metrics-annotation:4.1.20}} has the property \texttt{\path{version}} in different order~\footnote{\url{https://github.com/chains-project/reproducible-central/issues/17\#issuecomment-2576045821}}.

\textbf{Solution} The order of the properties should be fixed in the \texttt{\path{pom.properties}} file or the file should be removed from the JAR file as it is not required for the JAR file to run.

\subsection{Timestamp}
\label{appendix:pom-properties-timestamp}

While generating \texttt{\path{pom.properties}} file, Maven uses the current time to set the time of generation.

\textbf{Example} \texttt{\path{ch.qos.logback.db:logback-classic-db:1.2.11.1}} embeds timestamp in one of the comments in the \texttt{\path{pom.properties}} file.

\begin{lstlisting}
  #Generated by Apache Maven
  #Wed Apr 20 20:27:33 CEST 2022
  #Wed Jan 29 08:18:40 UTC 2025
\end{lstlisting}

\textbf{Solution} The timestamp should either be set to the value of \texttt{\path{SOURCE_DATE_EPOCH}} or removed from the \texttt{\path{pom.properties}} file.

\subsection{Eclipse properties}
\label{appendix:pom-properties-eclipse}

If a Maven module is built using, \texttt{\path{m2e}} --- Maven build tool for Eclipse IDE, it adds some properties to the \texttt{\path{pom.properties}} file other than the ones added by standalone Maven build tool.

\textbf{Example} Maven release \texttt{\path{org.spdx:spdx-maven-plugin:0.7.0}} has \texttt{\path{m2e.projectLocation}} and \texttt{\path{m2e.projectName}} properties in the \texttt{\path{pom.properties}} file.

\textbf{Solution} These attributes can be removed from the \texttt{\path{pom.properties}} file as they are not required for the JAR file to run.

\section{Unreproducible Software Bill of Materials}

\subsection{Removal of hash algorithms}
\label{appendix:sbom-hash-algo}

CycloneDX supports multiple hash algorithms to calculate the hash of components.
Our dataset shows that some rebuilds omits some of the hash algorithms.

\textbf{Example} \texttt{\path{ratis-3.1.1-cyclonedx.json}} artifact of Maven release \texttt{\path{org.apache.ratis:ratis:3.1.1}} has different hash algorithms in the CycloneDX SBOM.
The rebuild version of the SBOM omits the \texttt{\path{SHA3-384}}, \texttt{\path{SHA3-256}}, and \texttt{\path{SHA3-512}} hash algorithms.

\textbf{Solution} These hash algorithms are not present in JDK \texttt{\path{1.8.0_422}} which is used to rebuild the artifact in our dataset.
The solution is to use a JDK that has backported these hash algorithms.
This has already been solved by modifying the \texttt{\path{buildspec}} file to use Azul JDK~\footnote{https://github.com/jvm-repo-rebuild/reproducible-central/commit/ff0cc3037b7a1d0f009d4e4461c10f817c8658c7}.

\subsection{Addition or removal of \texttt{components}}
\label{appendix:sbom-components-add-remove}

SBOM records the components used to build the software.
In the context of Maven, this can either be a dependency or a Maven submodule.
Difference in components in SBOM mean that the dependencies are different across different builds and hence the build is unreproducible.

\textbf{Example} Maven release \texttt{\path{net.sourceforge.pmd:pmd:7.0.0}} has an artifact \texttt{\path{pmd-7.0.0-cyclonedx.json}} which shows that the \texttt{\path{components}} attribute is different across reference and rebuild versions of the SBOM.
The rebuild version documents two more Maven submodules which are \texttt{\path{net.sourceforge.pmd:pmd-cli:7.0.0}} and \texttt{\path{net.sourceforge.pmd:pmd-dist:7.0.0}}.

\textbf{Solution} \newc{The build script here\footnote{\url{https://github.com/pmd/pmd/blob/7979570d39909195e173de05450fb1acf8b3eec5/.ci/build.sh\#L81}} shows ad-hoc configuration of the build process.}
This should not be done as it hinders automation when checking for reproducibility.
The fix here is to stick to conventional builds.

\textbf{Example} Maven release \texttt{\path{io.dropwizard.metrics:metrics-core:4.1.32}} has a CycloneDX SBOM, \texttt{\path{metrics-core-4.1.32-cyclonedx.json}}, where all \texttt{\path{components}} are deleted except \texttt{\path{org.slf4j:slf4j-api:1.7.36}}.

\textbf{Solution} It is unclear why all components are deleted.
The original dataset does not have the same result~\footnote{\url{https://github.com/jvm-repo-rebuild/reproducible-central/blob/55c06c8c4a080e66267f573560166c42071b2814/content/io/dropwizard/metrics/metrics-parent-4.1.32.diffoscope}} asked here:\url{https://github.com/dropwizard/metrics/issues/4702}.
The solution that Reproducible Central uses is to ignore checking for generated SBOMs.

\subsection{Modification of \texttt{components}}
\label{appendix:sbom-components-modification}

Even if the components are the same across different builds, the content of the components should exactly match across different builds.

\textbf{Example} Maven release \texttt{\path{org.apache.bcel:bcel:6.8.1}} has an artifact \texttt{\path{bcel-6.8.1-cyclonedx.json}} where one of the components \texttt{\path{org.apache.commons:commons-lang3:3.14.0}} has different hashes.
SHA1 hash in reference versions is \texttt{\path{29a8e03}} while in the rebuild version is \texttt{\path{1ed4711}}.
The rebuild version's hash is also consistent with the one on Maven Central~\footnote{\url{https://repo1.maven.org/maven2/org/apache/commons/commons-lang3/3.14.0/commons-lang3-3.14.0.jar.sha1}}.
This happens because the local ~\texttt{\path{.m2}} folder has locally installed version which is referenced in the SBOM.
However, the rebuild version uses the version from Maven Central.

\textbf{Solution} The release command should clean the local Maven repository before building the artifact.
This can be done using \texttt{\path{purge-local-repository}} goal of \texttt{\path{maven-dependency-plugin}}~\footnote{https://maven.apache.org/plugins/maven-dependency-plugin/examples/purging-local-repository.html}.

\subsection{\texttt{metadata.timestamp}}
\label{appendix:sbom-timestamp}

An attribute in the CycloneDX SBOM is used to indicate the time when the SBOM is generated.
This can be different across different builds as the build time is different.

\textbf{Example} The CycloneDX SBOM, \texttt{\path{cyclonedx-core-java-7.3.2-cyclonedx.json}} of Maven release \texttt{\path{org.cyclonedx:cyclonedx-core-java:7.3.2}} has different \texttt{\path{metadata.timestamp}} attribute.

\textbf{Solution} It should respect the Maven property \texttt{\path{${project.build.outputTimestamp}}} to set the value of \texttt{\path{metadata.timestamp}} attribute.
Although this has been addressed in pull request~\footnote{https://github.com/CycloneDX/cyclonedx-core-java/pull/63} in release \texttt{\path{3.0.8}}, but it does not seem to resolve the issue until release \texttt{\path{8.0.0}}.
Perhaps, a better way to deal with this is to canonicalize the timestamp attribute.

\subsection{\texttt{serialNumber}}
\label{appendix:sbom-serial-number}

A unique serial number is assigned to an SBOM even if the content of the SBOM is the same~\footnote{\url{https://cyclonedx.org/docs/1.6/json/\#serialNumber}}.

\textbf{Example} Maven release \texttt{\path{org.cyclonedx:cyclonedx-maven-plugin:2.7.5}} produces \texttt{\path{cyclonedx-maven-plugin-2.7.5-cyclonedx.xml}} where difference in \texttt{\path{serialNumber}} attribute is observed.

\textbf{Solution} The \texttt{\path{serialNumber}} attribute should be removed before comparison as it is not relevant to the content of the SBOM.
Else, the generation should depend upon some fixed content of the SBOM to ensure the same serial number across different builds.
This has been addressed by computing \texttt{\path{serialNumber}} on the basis of groupId, artifactId, and version~\footnote{\url{https://github.com/CycloneDX/cyclonedx-maven-plugin/pull/425}}.

\subsection{External Reference}
\label{appendix:sbom-external-reference}

\newc{This attribute indicates external systems which are related to the software.
For example, version control system, issue tracking system, documentation, etc.}

\newc{\textbf{Example} Gradle release \texttt{\path{org.jreleaser:jreleaser-mattermost-java-sdk:1.6.0}} has an SBOM that only shows VCS reference for the components in the rebuilt version of the SBOM.}

\newc{\textbf{Solution} The external reference should be removed via canonicalization.}

\subsection{Description of components}
\label{appendix:sbom-description-components}

\newc{The description attribute is a free-form text that can be used to describe the component.}

\newc{\textbf{Example} Gradle release \texttt{\path{org.jreleaser:jreleaser-nexus2-java-sdk:1.6.0}} has an SBOM that only shows description for the components in the rebuilt version of the SBOM.}

\newc{\textbf{Solution} The description of the components should be removed via canonicalization.}

\subsection{License}
\label{appendix:sbom-license}

\newc{The difference in licence appears in Software Package Data Exchange (SPDX) format.
SPDX is a standard to document the licenses of the components used to build the software.
Apart from timestamps, it has a non-deterministic attribute \texttt{\path{licenseListVersion}} which can cause unreproducibility.
\texttt{\path{licenseListVersion}} attribute is used to indicate the version of the SPDX license list used to document the licenses of the components.
This list is updated externally and can vary with time.}

\newc{\textbf{Example}
Maven release \texttt{\path{org.apache.commons:commons-parent:60}} has an artifact \texttt{\path{commons-parent-60.spdx.json}} where value of timestamp and \texttt{\path{licenseListVersion}} attribute is different across different builds.}

\newc{\textbf{Example}
Gradle release \texttt{\path{org.jreleaser:jreleaser-mattermost-java-sdk:1.6.0}} has an SBOM that captures the license URLs for more components.}

\newc{\textbf{Solution} License attribute should be removed before comparison as they are non-deterministic.}

\section{Unreproducible Filesystem}

\subsection{Files are removed or added}
\label{appendix:files-removed-added}

If archives have different number of files, then the archive cannot be reproducible as the content of the archive is different.

\textbf{Example} Maven release \texttt{\path{commons-daemon:commons-daemon:1.4.0}} has an unreproducible artifact \texttt{\path{commons-daemon-1.4.0-bin-windows.zip}} which does not contain three Windows executable binary in the rebuild version of the archive.
There is custom non-Maven command that has to be executed during the build process to generate and place these binaries under the build folder.
Since the command is not documented in the \texttt{\path{buildspec}} file, the rebuild version of the archive does not contain these binaries.

\textbf{Solution} The build process should be fixed to incorporate generation of other binaries along with the Maven build command.
If this is not possible, the rebuilder should document the steps for generating the binaries in the \texttt{\path{buildspec}} file.

\textbf{Example} Maven release \texttt{\path{org.apache.rat:apache-rat-project:0.16.1}} has an unreproducible artifact \texttt{\path{apache-rat-0.16.1-src.tar.bz2}} which has an empty directory \texttt{\path{apache-rat-0.16.1/apache-rat-api/}} compressed in the archive.
This directory has been deleted in this version\footnote{\url{https://github.com/apache/creadur-rat/blob/c5a31fecc3b6d3697e20cb867e82c55decf969be/RELEASE-NOTES.txt\#L22}}.
However, it existed in the earlier versions and while checking out only the files are deleted but directory is kept as git does not track directories.

\textbf{Solution} The rebuilder should remove the empty directories from the source code before building the artifact.
\texttt{\path{git clean -df}} can be used to remove the empty directories from the source code.

\textbf{Example} Maven release \texttt{\path{org.apache.activemq:apache-artemis:2.28.0}} has source archive, \texttt{\path{apache-artemis-2.28.0-source-release.tar.gz}}, as its artifact.
It has pushed the node modules in the archive which are not present in the rebuild version.

\textbf{Solution} The builder should avoid pushing the node modules in the source archive and should only push the files that are tracked by the version control system.
Ideally, the node modules are always untracked to reduce the size of git repository.

\subsection{File permissions}
\label{appendix:file-permissions}

The file in the archive can be created with different permissions across different builds.
\texttt{\path{umask}} utility controls the permission of newly created files and directories and it can be different across different environments.

\textbf{Example} Maven release \texttt{\path{io.github.albertus82:unexepack:0.2.1}} has an unreproducible artifact \texttt{\path{unexepack-0.2.1-bin.zip}} which has different file permissions for the file \texttt{\path{unexepack-0.2.1/unexepack.jar}} in the rebuild version of the archive.

\textbf{Solution} The rebuilder can either set the file permissions to a fixed value or set \texttt{\path{umask}} in the rebuilder script.

\subsection{File ownership}
\label{appendix:file-ownership}

The file in the archive can be created with different ownership across different builds.

\textbf{Example} Maven release \texttt{\path{org.apache.accumulo:accumulo:1.10.2}} has an artifact \texttt{\path{accumulo-1.10.2-src.tar.gz}}.
The file ownership has changed from \texttt{\path{christopher}} to \texttt{\path{aman}}.

\textbf{Solution} File ownership should be set to a fixed value by the rebuilder from all artifacts before comparison.

\subsection{File timestamps}
\label{appendix:file-timestamps}

\textbf{Example} \texttt{\path{org.apache.zookeeper:parent:3.8.1}} upon rebuilding produces \texttt{\path{parent-3.8.1.tar.gz}} where the timestamp of the files in the tar ball is set to the time of rebuild.

\textbf{Solution} The timestamp can either be set to a fixed value or stripped from the file by rebuilder before comparison.

\subsection{File type}
\label{appendix:file-type}

The file type of archive are different across different builds.
This causes the size of the archives to be different as different compression algorithms could be used.

\textbf{Example} Consider Maven release \texttt{\path{io.github.git-commit-id:git-commit-id-maven-plugin:6.0.0}} which has an unreproducible artifact \texttt{\path{git-commit-id-maven-plugin-6.0.0-sources.jar}}.
It shows that the type of the file changes from \texttt{\path{Zip archive data, at least v1.0 to extract, compression method=store}} to \texttt{\path{Java archive data (JAR)}}\footnote{Seems different from RC \url{https://github.com/jvm-repo-rebuild/reproducible-central/blob/329fd7e4d7721cf7ea85fe203fadfb884be13fa9/content/io/github/git-commit-id/git-commit-id-maven-plugin-6.0.0.diffoscope\#L129-L130}}.
This seems to happen because outdated plugins are used to generated the source archive as it has attributes like \texttt{\path{Built-By}} in the MANIFEST.MF file which has been removed in the newer version of \texttt{\path{maven-archiver}} plugin~\footnote{\url{https://issues.apache.org/jira/browse/MSHARED-799}}.

\textbf{Solution} The builder should always use the specified version of the plugins to generate the archives.
To help with this, the local repository can be cleaned before building the artifact using \texttt{\path{purge-local-repository}}~\footnote{\url{https://maven.apache.org/plugins/maven-dependency-plugin/examples/purging-local-repository.html}}.

\textbf{Example} 

\subsection{File size}
\label{appendix:file-size}

\textbf{Example} \texttt{\path{org.apache.maven:apache-maven:3.8.1}} has the executable \texttt{\path{mvn}} whose size is different across different builds.
It changes from \texttt{\path{5741}} bytes to \texttt{\path{5940}} bytes because of different new lines in both executables.

\textbf{Solution} The rebuilder should replace the new lines from all files to either \texttt{\path{LF}} or \texttt{\path{CRLF}} before comparison based on the operating system of the builder.
Builder should not replace the new lines as it could break behavior of the executable on different platforms~\footnote{\url{https://cwiki.apache.org/confluence/pages/viewpage.action?pageId=74682318}}.

\subsection{File names}
\label{appendix:file-names}

\textbf{Example} \texttt{\path{org.apache.jena:jena-permissions:4.3.2}} has an unreproducible artifact \texttt{\path{jena-permissions-4.3.2-sources.jar}} where the file name is different in the rebuild version of the archive, but the content is identical.

\textbf{Solution} The builder should not embed system specific paths in any artifact~\cite{andersson_geth_2024}.
Go ecosystem has a tool \texttt{\path{trimpath}} which removes the system specific paths.

\subsection{File paths embedded}
\label{appendix:file-embedded}

If absolute file paths are embedded in the archive, then the build is unreproducible.
In our dataset, we observe that the absolute file paths are embedded in JVM bytecode and configuration files.

\textbf{Example} Maven release \texttt{\path{org.apache.cxf.fediz:common:1.6.1}} has an artifact \texttt{\path{common-1.6.1.jar}} which has a file path embedded in the JVM bytecode.

\textbf{Example} Maven release \texttt{\path{org.owasp:dependency-check-cli:7.0.4}} embeds absolute paths in \texttt{\path{dependencycheck-cache.properties}}.

\textbf{Solution} The builder should not embed absolute paths in the configuration files and stick to relative paths.
\section{Unreproducible JVM bytecode}

\subsection{Debug information}
\label{appendix:jvm-debug-info}

This corresponds to the information in the JVM bytecode that is used for debugging the Java program.
There are three types of debug information in the JVM bytecode --- \texttt{\path{SourceFile}}, \texttt{\path{LineNumberTable}}, and \texttt{\path{LocalVariableTable}}.

\textbf{Example} In release \texttt{\path{io.dropwizard.metrics:metrics-httpclient:4.1.27}}, the artifact \texttt{\path{metrics-httpclient-4.1.27.jar}} shows an extra line in LineNumberTable which corresponds to the \texttt{\path{return}} JVM bytecode instruction for a void method.
This could be because the precise version and vendor of JDK is not used.

\textbf{Solution} The rebuilder should remove the information related to debugging from the JVM bytecode before comparison.

\subsection{Optimizations or de-optimizations}
\label{appendix:jvm-optimizations}

The JVM bytecode can be optimized or de-optimized across different builds if the JDK version used to build the artifact is different.
Newer versions of JDK can have better optimizations while older versions can have less performant code.
We have found the following types of differences in the JVM bytecode across different builds:
\begin{itemize}
  \item try-with-resource is simplified with less \texttt{\path{goto}} instructions and \texttt{\path{close()}} invocations.
  Example: \texttt{\path{io.dropwizard.metrics:metrics-servlets:4.1.20}}.
  \item return type in \texttt{\path{invokevirtual}} change from \texttt{\path{java/nio/Buffer}} to \texttt{\path{java/nio/ByteBuffer}}. Example: \texttt{\path{io.dropwizard.metrics:metrics-collectd:4.1.20}}.
  \item \texttt{\path{invokeinterface}} is changed to \texttt{\path{invokevirtual}}. Example: \texttt{\path{com.github.ldapchai:ldapchai:0.8.6}}.
  \item implementation \texttt{\path{finally}} is changed from using subroutines (\texttt{\path{jsr}} and \texttt{\path{ret}}) to using more inline code. Example: \texttt{\path{net.bytebuddy:byte-buddy-dep:1.14.5}}.
  \item \texttt{\path{static}} and \texttt{\path{final}} modifiers are added or removed from anonymous classes. Example: \texttt{\path{dev.langchain4j:langchain4j-core:0.26.0}}.
  \item invocations to \texttt{\path{Objects.requireNonNull}} are added in rebuild version of the JVM bytecode. Example: \texttt{\path{dev.langchain4j:langchain4j-core:0.26.0}}.
  \item \texttt{\path{checkcast}} instructions are reduced. Example: \texttt{\path{org.apache.sling::org.apache.sling.servlets.resolver:2.8.2}}.
  \item implementation of nested enums differ. This is due to a commit in OpenJDK 22~\footnote{\url{https://github.com/openjdk/jdk/pull/10797}} which changes how switch map tables are generated.
  \item synthetic method, \texttt{\path{values()}}, has bytecode under a separate method in reference version and under the static initializers in rebuild version~\footnote{\url{https://bugs.openjdk.org/browse/JDK-8241798}}. Example: \texttt{\path{com.google.guava:guava:32.0.0-jre}}
  \item synthetic field for inner class is added in rebuild version~\footnote{https://bugs.openjdk.org/browse/JDK-8271717}. Example: \texttt{\path{com.google.guava:guava:32.0.0-jre}}
  \item local variables in \texttt{\path{LocalVariableTable}} are in the scope for a more number of instructions. Example: \texttt{\path{org.apache.helix:helix-common:1.0.3}}.
  \item \texttt{\path{ldc_w}} instructions are replaced with \texttt{\path{ldc}} instructions. Example: \texttt{\path{org.codehaus.mojo:jaxb2-maven-plugin:3.1.0}}.
  \item strings are concatenated which reduces the number of \texttt{\path{ldc}} instructions. Example: \texttt{\path{org.apache.sling.event:org.apache.sling.event:4.2.18}}.
  \item \texttt{\path{StackMapTable}} is modified. \texttt{\path{org.apache.sling:org.apache.sling.servlets.resolver:2.8.2}}
\end{itemize}

\textbf{Solution} The rebuilder should use the precise version of JDK to build the artifact as the reference version.
This is a non-trivial task as the exact version of JDK is not always mentioned and there are a myriad of JDK versions and vendors available.
As an example, we found that building \texttt{\path{com.github.ldapchai:ldapchai:0.8.6}} with JDK 22 solves all differences related to JVM bytecode.

\subsection{Refactoring}
\label{appendix:jvm-refactoring}

These changes simply change the structure of the JVM bytecode without altering the semantics or performance.
Changes in order of sections in JVM bytecode is an example of refactoring.
The JVM bytecode contains sections that are order independent, but they can differ across different builds and fail the build reproducibility check.
In our dataset, we observe that the order of methods, fields, static initializers, entries in the constant pool, array type values in annotation, method declaration in interfaces, and order of inner classes can differ across different builds.

\textbf{Example} In Maven release \texttt{\path{org.apache.activemq:activemq-runtime-config:5.16.4}}, the artifact \texttt{\path{activemq-runtime-config-5.16.4.jar}} uses \texttt{\path{jakarta.xml.bind.annotation.XmlElementRefs}} which takes an array of \texttt{\path{XmlElementRef}} annotations.
The order of these values in the array is different across different builds.

\textbf{Solution} The rebuilder should sort the order independent sections of the JVM bytecode before comparison.

Apart from order, we observe two types of refactoring that did not change the order.
\begin{itemize}
\item \texttt{\path{private}} modifier is removed from a synthetic field in an anonymous class in artifact \texttt{\path{antisamy-1.7.3.jar}}. We consider this as a refactoring as fields in anonymous classes are implicitly \texttt{\path{private}}.
\item The positive and negative branches of an \texttt{\path{if}} statement are swapped in the JVM bytecode of Maven release \texttt{\path{eu.maveniverse.maven.mima:2.4.2:standalone-static-uber}}
\end{itemize}

\textbf{Solution} The rebuilder should canonicalize the JVM bytecode before comparison.

\subsection{Embedded data}
\label{appendix:embedded-data}

Absolute file paths, timestamp, java version, project version and other environment related variables are embedded in the JVM bytecode.

\textbf{Example} Maven release \texttt{\path{org.apache.hive:hive-exec:4.0.0-alpha-2}} embeds git branch name, user name, timestamp, and git repository path in \texttt{\path{hive-exec-4.0.0-alpha-2.jar}} which makes it unreproducible.

\textbf{Solution} The builder should embed such environment related variables in the JVM bytecode.

\subsection{Build time generated/modified code}
\label{appendix:build-time-generated-code}

The Java compiler produces synthetic code during build in order to handle access to members~\cite{rimenti_synthetic_2018}.
The compiler also adds names to lambda functions~\cite{evans_scenes_2020} and inner classes~\cite{maverik_why_2016} which are not present in the source code.
In addition to these capabilities, Maven provides plugins to generate or modify code during the build process.
All of this generated code can be different across different builds and hence the build is unreproducible.

\textbf{Example} Maven release \texttt{\path{io.github.chains-project:maven-lockfile-github-action:3.4.0}} generates a lot of code during the build process.
The code is generated by two dependencies of maven-lockfile, \texttt{\path{o.quarkiverse.githubaction:quarkus-github-action:2.0.1}} and \texttt{\path{io.quarkus:quarkus-arc:3.1.1.Final}} in order to make the application compatible with GitHub actions.
The generated code differs in order of fields, different constant values, and embedded system properties.
Another example is \texttt{\path{org.apache.hive:kafka-handler:4.0.0-alpha-2}} which uses \texttt{\path{maven-shade-plugin}} to rename package names from \texttt{\path{kafkaesqueesque}} to \texttt{\path{kafkaesque}}.

\textbf{Example} Maven release \texttt{\path{org.apache.helix:zookeeper-api:1.0.3}} has an artifact \texttt{\path{zookeeper-api-1.0.3.jar}}.
The lambda function changes from \texttt{\path{lambda$parseRoutingData$19}} to \texttt{\path{lambda$parseRoutingData$1}}.

\textbf{Example} Maven release \texttt{\path{org.apache.hive:hive-parser:4.0.0-alpha-2}} has an artifact \texttt{\path{hive-parser-4.0.0-alpha-2-sources.jar}} that has sources generated by ANTLR3 Maven plugin.
The generated code has comments where order of values differs across builds.
This issue is fixed by sorting the values in the comments~\footnote{\url{https://github.com/antlr/antlr4/pull/3809}}.

\textbf{Example} Maven release \texttt{\path{org.apache.shiro:shiro-aspectj:1.11.0}} produces an artifact where AspectJ metadata is embedded in the JVM bytecode.
The metadata does not exist in the reference build.
This issue has been fixed by forcing the compilation of AspectJ classes~\footnote{\url{https://github.com/apache/shiro/commit/2d92460cd8b7a621be0b82725b7a5cd6952734c1}}.

\newc{\textbf{Example} Gradle release \texttt{\path{io.opentelemetry.instrumentation:opentelemetry-instrumentation-api:2.4.0}} has an artifact \texttt{\path{opentelemetry-javaagent-2.4.0.jar.diffoscope.json}} which records MethodParameters attribute only the in reference version of the bytecode to allow reflective access to the name of the arguments.}

\textbf{Example} Maven release \texttt{io.fabric8:kubernetes-model\\-apiextensions:6.4.0} contains build time generated code where the order of if-else blocks is different across different builds.
We do not classify them under refactoring as the order of if-else blocks is relevant to the semantics of the code.

\textbf{Solution} The generated code should be deterministic across different builds.
This calls for fixes in the Java compiler or the Maven plugins that generate code during the build process.

\section{Unreproducible Timestamps}

\subsection{Properties file}
\label{appendix:timestamp-properties}

The properties files in a JAR file stores configuration related to the project.
This can be either a file manually created by the developer or generated by the build tools.
In our dataset, we observe that the timestamps in the properties files are different across different builds.

\textbf{Example} Maven release \texttt{\path{com.adobe.acs:acs-aem-commons-bundle:5.2.0}} has files embedded in the JAR file that have different timestamps of generations across different builds.

\textbf{Solution} The plugin that generates the properties file should respect the Maven property \texttt\texttt{\path{${project.build.outputTimestamp}}} to set the value of the timestamp in the properties file.

\subsection{Documentation}
\label{appendix:timestamp-doc}

The generated documentation can have timestamps embedded in it.

\textbf{Example} Maven release \texttt{\path{org.apache.synapse:synapse-documentation:3.0.2}} has an artifact \texttt{\path{synapse-documentation-3.0.2.jar}} which has a timestamp embedded in the documentation.
\texttt{\path{org.apache.karaf:manual:4.4.0}} is another example.

\textbf{Solution} The plugin that generates the documentation should respect the Maven property \texttt{\path{${project.build.outputTimestamp}}} to set the value of the timestamp in the documentation.

\subsection{Shell scripts}
\label{appendix:timestamp-shell}

\textbf{Example} Maven release \texttt{\path{org.apache.sling:org.apache.sling.feature.cpconverter:1.3.4}} has an artifact \texttt{\path{org.apache.sling.feature.cpconverter-1.3.4.tar.gz}} that packs a shell script \texttt{\path{bin/cp2feature}}.
The script takes build timestamp as one of the arguments and the timestamp is different across different builds.

\textbf{Solution} The shell script should respect the Maven property \texttt{\path{${project.build.outputTimestamp}}} or the standard \texttt{\path{SOURCE_DATE_EPOCH}} environment variable to set the value of the timestamp in the shell script.

\subsection{Binaries}
\label{appendix:timestamp-binaries}

\textbf{Example} Maven release \texttt{\path{io.github.albertus82:unexepack:0.2.1}} embeds a Windows executable that embeds the timestamp.

\textbf{Solution} The build tool for binary should respect \texttt{\path{SOURCE_DATE_EPOCH}} environment variable to set the value of the timestamp in the binary.

\subsection{SBOM}
\label{appendix:timestamp-sbom}

Unreproducibility due to attribute \texttt{\path{timestamp}} in the SBOM as explained in \cref{appendix:sbom-timestamp}.

\subsection{JVM bytecode}
\label{appendix:timestamp-jvm}

JVM bytecode can have timestamps embedded as values of variables.

\textbf{Example} package-info.class in \texttt{\path{4.0.0-alpha-2:hive-standalone-metastore-server:4.0.0-alpha-2}} has timestamp embedded in an annotation.

\textbf{Solution} This should respect the Maven property \texttt{\path{${project.build.outputTimestamp}}} to set the value of the timestamp in the JVM bytecode.

\subsection{File timestamps}
\label{appendix:timestamp-files}

Files have timestamp for creation or modification time as shown in \cref{appendix:file-timestamps}.

\subsection{MANIFEST}
\label{appendix:timestamp-manifest}

Attribute \texttt{\path{Bnd-LastModified}} tends to be non-deterministic as documented in \cref{appendix:mf-bnd-lastmodified}.

\subsection{NOTICE}
\label{appendix:timestamp-notice}

\textbf{Example} Maven release \texttt{\path{org.apache:apache:24}} has a sources Jar where the year of build is embedded in the \texttt{\path{apache-24/NOTICE}} file.

\textbf{Solution} The plugin that generates the NOTICE file should respect the Maven property \texttt{\path{${project.build.outputTimestamp}}} to set the value of the timestamp in the NOTICE file.

\subsection{Servlets}
\label{appendix:timestamp-servlets}

\textbf{Example} Maven release \texttt{\path{org.apache.nifi:nifi-web-ui:1.27.0}} has embedded timestamps.
These timestamps are created by Jasper which compiles JSP files to servlets.

\textbf{Solution} The plugin that generates the servlets should respect the Maven property \texttt{\path{${project.build.outputTimestamp}}} to set the value of the timestamp in the servlets.

\section{Unreproducible Order}

\subsection{Ordering in JVM bytecode}
\label{appendix:order-jvm}

The JVM bytecode contains sections that are order independent, but they can differ across different builds and fail the build reproducibility check.
In our dataset, we observe that the order of methods, fields, static initializers, entries in the constant pool, array type values in annotation, method declaration in interfaces, and order of inner classes can differ across different builds.

\textbf{Example} In Maven release \texttt{\path{org.apache.activemq:activemq-runtime-config:5.16.4}}, the artifact \texttt{\path{activemq-runtime-config-5.16.4.jar}} uses \texttt{\path{jakarta.xml.bind.annotation.XmlElementRefs}} which takes an array of \texttt{\path{XmlElementRef}} annotations.
The order of these values in the array is different across different builds.

\textbf{Solution} The rebuilder should sort the order independent sections of the JVM bytecode before comparison

\cref{appendix:jvm-refactoring}

\subsection{pom.properties}
\label{appendix:order-pom}

The attributes in the \texttt{\path{pom.properties}} file are not in the same order across different builds.

\textbf{Example} \texttt{\path{io.dropwizard.metrics:metrics-jcache:4.1.20}} has an artifact \texttt{\path{metrics-jcache-4.1.20.jar}} which has a \texttt{\path{version}} attribute that appears at line 1 in the reference version and at line 3 in the rebuild version.
\texttt{\path{io.takari.maven:takari-smart-builder:1.0.0}} is another example where order of \texttt{\path{artifactId}} is different.

\textbf{Solution} The rebuilder should sort the attributes in the \texttt{\path{pom.properties}} file before comparison.
However, this file is most commonly generated by \texttt{\path{maven-archiver-plugin}} and this ordering issue has been fixed~\footnote{https://github.com/apache/maven-archiver/commit/763a940540eefad74f9ba73cb5eed288dc4e639d}.

\subsection{XML files}
\label{appendix:order-xml}

\textbf{Example} Order of \texttt{\path{remoteResources}} in file generated by \texttt{\path{maven-remote-resources-plugin}} is different across different builds.
This is seen in \texttt{\path{org.apache.axis2:axis2-resource-bundle:1.8.2}}.

\textbf{Solution} The rebuilder should sort the order of elements in the XML files before comparison.
This won't affect the future releases as this unstable order has been fixed in \texttt{\path{3.3.0}} version of \texttt{\path{maven-remote-resources-plugin}}~\footnote{\url{https://github.com/apache/maven-remote-resources-plugin/pull/74}}.

\textbf{Example} \texttt{\path{components.xml}} generated by \texttt{\path{plexus-containers}} has different order of elements across different builds.

\textbf{Solution} The rebuilder should sort the order of elements in the XML files before comparison.
This has been fixed recent releases of \texttt{\path{plexus-containers}}~\footnote{\url{https://github.com/codehaus-plexus/plexus-containers/issues/8}}.

\subsection{JMH benchmark}
\label{appendix:order-jmh}

JMH is a microbenchmarking framework for Java.

\textbf{Example} \texttt{\path{org.apache.hive:hive-metastore-benchmarks:4.0.0-alpha-2}} publishes output from JMH benchmarks in the JAR file.
Even though the contents remain the same, the order of results is different across different builds.

\textbf{Solution} Builder should not publish the output from JMH benchmarks in the JAR file.
It is a metric that is used to improve the performance of the code and should not be part of the released artifact.

\subsection{Files in archive}
\label{appendix:order-files-archive}

Jar file is a zip file and the order of files can be in any order.

\textbf{Example} Order of files in \texttt{\path{org.apache.synapse:synapse-package-archetype:3.0.2}} do not stay the same across builds.

\textbf{Solution} The order of zip files can be canonicalized by rebuilder before comparison.
For future releases, this has been fixed~\footnote{\url{https://issues.apache.org/jira/browse/MJAR-263}}~\footnote{\url{https://issues.apache.org/jira/browse/MSHADE-347}}.

\subsection{Generated files}
\label{appendix:order-generated-files}

\textbf{Example} Maven release \texttt{\path{org.apache.hive:hive-parser:4.0.0-alpha-2}} generates a file using ANTLR3 Maven plugin which creates a different order of values in the comments across different builds.

\textbf{Solution} The plugin that generates the file should sort the order of values in the comments before comparison.
ANTLR3 is not maintained anymore but this issue has been fixed in ANTLR4~\footnote{\url{https://github.com/antlr/antlr4/pull/3809}}.

\textbf{Example} \texttt{\path{net.sourceforge.pmd:pmd-cli:7.0.0-rc3}} generates a shell completion script which has different order of flags across different builds.

\textbf{Solution} The plugin that generates the file should sort the order of flags in the shell completion script before comparison.
This has been fixed in later version of \texttt{\path{pmd-cli}}.

\textbf{Example} Apache Tomcat JspC creates \texttt{\path{web.xml}} file with Java Server Pages configuration.
We observe that the order declaration is different across builds in Maven release \texttt{\path{org.apache.hive:hive-service:4.0.0-alpha-2}}.

\textbf{Solution} The plugin that generates the file should sort the order of declarations in the \texttt{\path{web.xml}} file before comparison.

\subsection{Attribute values in MANIFEST}
\label{appendix:order-mf}

Some values are comma separated lists whose order can differ as shown in \cref{appendix:mf-order}.

\section{Mysteries}
\label{appendix:mysteries}

\begin{itemize}
  \item Target property is changed from 8 to 1.8 and JVM option removed in dependency-reduced-pom.xml \url{https://github.com/chains-project/reproducible-central/issues/20#issuecomment-2581417522}
  \item Change in generated documentation \url{https://github.com/chains-project/reproducible-central/issues/20#issuecomment-2582710398}
  \item URL to Nexus repository is replaced with a new one in \texttt{\path{io.dropwizard.metrics:metrics-parent:4.2.6}} \url{https://github.com/chains-project/reproducible-central/issues/20#issuecomment-2583498935}.
  \item Configuration for \texttt{\path{maven-failsafe-plugin}} is deleted in \texttt{\path{org.apache.maven.plugins:maven-gpg-plugin:3.1.0}}. \url{https://github.com/chains-project/reproducible-central/issues/20#issuecomment-2583617364}
  \item URL change in license headers \url{org.apache.drill:distribution:1.20.0}. \url{https://github.com/chains-project/reproducible-central/issues/20#issuecomment-2583664806}
  \item Change in \texttt{\path{registry.json}} in \texttt{\path{org.apache.drill:drill-common:1.21.1}}. \url{https://github.com/chains-project/reproducible-central/issues/20#issuecomment-2583691367}
  \item Changes in META-INF files \url{https://github.com/chains-project/reproducible-central/issues/20#issuecomment-2583778244}
  \item Differences in encoding \url{https://github.com/chains-project/reproducible-central/issues/20#issuecomment-2583612194}
  \item Changes in POM.xml \url{https://github.com/chains-project/reproducible-central/issues/20#issuecomment-2584322171}
  \item \texttt{\path{groupId}} under \texttt{\path{exclusion}} varies. \url{https://github.com/chains-project/reproducible-central/issues/17#issuecomment-2576071584}
  \item \texttt{\path{org.apache.turbine:turbine-webapp-6.0:3.0.0}} changes \texttt{\path{2011-1969}} to \texttt{\path{2011-1970}} in copyright header.
  \item Content of NOTICE changes in \texttt{\path{org.apache.paimon:paimon-bundle:0.9.0}}.
  \item Order of flags of commands is flipped  autocompletion script. Maven release \texttt{\path{net.sourceforge.pmd:pmd-cli:7.0.0-rc3}}. Artifact \texttt{\path{pmd-cli-7.0.0-rc3-completion.sh}}
  \item Trailing slash is missing from one of the URLs in properties file~\url{https://github.com/chains-project/reproducible-central/issues/8#issuecomment-2566465525}.
  \item Interface is changed from \texttt{\path{Writable}} to \texttt{\path{Observable}} in \texttt{\path{org.apache.isis:isis-core-metamodel:2.0.0-M7}} \footnote{\url{https://github.com/chains-project/reproducible-central/issues/6\#issuecomment-2734000008}}
  \item Extra \texttt{\path{Recipe}} is returned in the list in \texttt{\path{tech.picnic.error-prone-support:error-prone-contrib:0.15.0}} ~\footnote{\url{https://github.com/PicnicSupermarket/error-prone-support/issues/1595}}
  \item The name of method parameter of \texttt{\path{java.lang.Thread}} is changed from \texttt{\path{arg0}} to \texttt{\path{name}} in \texttt{\path{dubbo-2.7.12.jar}}~\footnote{\url{https://github.com/chains-project/reproducible-central/issues/6\#issuecomment-2734158834}}
  \item Change in project.build.outputTimestamp \url{https://github.com/chains-project/reproducible-central/issues/18#issuecomment-2581344963}
  \item  Maven release \texttt{\path{org.mybatis:mybatis:3.5.16}} has an artifact \texttt{\path{mybatis-3.5.16.jar}} which has different \texttt{\path{X-Compile-Release-JDK}} attribute in the MANIFEST.MF file.
  It changes from 8 to 16. This is strange because source and target JDK are 16. \texttt{\path{X-Compile-Source-JDK}} and \texttt{\path{X-Compile-Target-JDK}} are contradicting \texttt{\path{X-Compile-Release-JDK}} attribute.
\end{itemize}

\section{Optimizations not canonicalized by \jNorm}
\label{appendix:jnorm-optimizations-failure}

The first case is how implementation enums differs between JDK 17 and JDK 22 compiled artifacts.
JDK 17 creates a lookup table to store the values of the enum while JDK 22 simply stores it based on the order of the enum values~\footnote{\url{https://github.com/openjdk/jdk/pull/10797}}.
We observe this difference in \texttt{\path{com.github.ldapchai:ldapchai:0.8.6}} as the reference version is built with JDK 22 and the rebuilt version is built with JDK 17.

% ldapchai also has the same issue (https://github.com/ldapchai/ldapchai/issues/32)
The second case is the usage of invokevirtual in JDK 17 and invokeinterface in JDK 21 bytecode instructions to execute \texttt{\path{toString}}, \texttt{\path{equals}}, and \texttt{\path{getClass}} methods.
They differ in the time the method is resolved.
The invokeinterface instruction resolves the method at runtime while invokevirtual resolves the method at compile time.
We observed this difference in our dataset in \texttt{\path{dev.langchain4j:langchain4j:0.28.0}} when \texttt{\path{toString}} method is invoked on instance of class.

The third case that \jNorm is unable to canonicalize the implementation of string concatenation when its operands require a cast to String.
For example, \texttt{\path{"Error determining JDBC type for column " + column + ".  Cause: " + e}} is a string concatenation expressions in artifact \texttt{\path{mybatis-3.5.16.jar}}.
\jNorm identifies that there are two ways to implement string concatenation~\cite{shipilev_jep_2024} --- using \texttt{\path{StringBuilder}} (JDK < 9) and using \texttt{\path{invokedynamic}} (JDK $\geq$ 9) instruction and canonicalizes the expression to the second one.
However, it fails to canonicalize the string conversion of \texttt{\path{e}} which is an instance of \texttt{\path{java.sql.SQLException}} before concatenation using \texttt{\path{invokedynamic}}.

Implementation of try and try-with-resource statements and finally clauses also differ across Java versions and is also not canonicalized by \jNorm.
For example, reference and rebuild artifacts of \texttt{\path{io.dropwizard.metrics:4.1.20:metrics-servlets-4.1.20}} are compiled with JDK 8 and JDK 11 respectively.
The rebuild artifact is optimized in the way it handles closing of resource.
Another example is \texttt{\path{net.bytebuddy:1.14.5:byte-buddy-1.14.5.jar}} that uses deprecated bytecode instructions \texttt{\path{jsr}} and \texttt{\path{ret}} to implement closing of resource.

\jNorm is unable to handle differences in how outer class reference is handled in the bytecode.
A synthetic field called \texttt{\path{this$0}} is created in the bytecode to refer to the outer class from inner class.
The reference version of \texttt{\path{com.google.guava:guava:32.0.1-jre}} is built using custom JDK which omits this synthetic field~\footnote{\url{https://bugs.openjdk.org/browse/JDK-8271717}} as it is not needed.
We also observe that patch versions of JDK 17 handle this differently.
They refer to the outer class reference using either \texttt{\path{this.this$0}} or \texttt{\path{this$0}}.
The former adds another \texttt{\path{getfield}} instruction to the bytecode.
\texttt{\path{psi-probe-tomcat85-3.7.0-tests.jar}} is an example of artifact that contains this difference.

\end{document}